\documentclass{article}

\usepackage{authblk}
\usepackage[utf8]{inputenc}
\usepackage[left=2cm,right=2cm,top=1.5cm,bottom=1.5cm]{geometry}
\usepackage{amsmath}
\usepackage{amssymb}
\usepackage{graphicx}
\usepackage{subfig}
\usepackage{mathtools}
\usepackage{mathpazo}
\usepackage[mathpazo]{flexisym}
\usepackage{breqn}
\usepackage{xcolor}
\usepackage{hyperref}
\usepackage{soul}
\hypersetup{
    colorlinks=true,
    linkcolor=blue,
    filecolor=magenta,      
    urlcolor=cyan,
    citecolor=teal
    }
\usepackage{comment}
\graphicspath{{figs/}}

\usepackage[style=numeric-comp,style=nature,url=true]{biblatex}
\DeclareMathOperator{\sech}{sech}

\title{ Radial kinks in a Schwarzschild-like geometry}
\author[1]{Jean-Guy Caputo}
\author[2]{Tomasz Dobrowolski}
\author[3]{Jacek Gatlik}
\author[4]{Panayotis G. Kevrekidis}
\affil[1]{\textit{Laboratoire de Math\'{e}matiques, INSA de Rouen, Avenue de l'Universit\'{e}, 76801 Saint-Etienne du Rouvray, France}}
\affil[2]{\textit{Department of Computer Physics and Quantum Computing, University of the National Education Commission in Krakow, Podchor\c{a}\.zych  2, 30-084 Cracow, Poland}}
\affil[3]{\textit{AGH University of Krakow, Faculty of Physics and Applied Computer Science, 30-059 Krakow, Poland}}
\affil[4]{\textit{Department of Mathematics and Statistics, University of Massachusetts, Amherst, Massachusetts 01003-4515, USA}}
\date{\today}

\begin{document}
\maketitle

\begin{abstract}
We study the propagation of a domain wall (kink) of the $\phi^4$ model in a radially symmetric environment defined by a gravity source. This
source introduces a Schwarzschild-like geometry. We introduce an effective model that accurately describes the dynamics of the kink center. This description works well even outside the perturbation region, i.e., even for large masses of the gravitating object.
We observed that such a spherical domain wall surrounding a star-type object inevitably 
``collapses'', i.e., shrinks in radius towards
the origin and offer an
understanding of the latter phenomenology.
The relevant analysis is presented for
a circular domain wall and a spherical one.
\end{abstract} \hspace{10pt}

\section{Introduction}
Topological defects are field solutions that play a significant role in many branches of science. The best known solutions of this type are domain walls, vortices, monopoles and textures \cite{Coleman1977, Vilenkin2000,mantonsut}. Solutions of this type describe real physical objects. Examples of such objects are vortices in superfluid quantum liquids (Helium-3 and Helium-4), magnetic flux tubes piercing superconductors of the II - type and domain walls present in ferromagnets and ferroelectrics \cite{pismen,Volovik2002, Bunkov2013, Jeff2000, Pouget1984, Kjems1978}. Various defects are also observed in liquid crystals \cite{Zannoni2002}. In addition to condensed matter physics many researchers speculate about possibility of their significance for the astrophysics and cosmology \cite{Kibble1976, Larsson1997, Dai2022}. 
In the cosmological background, topological defects may play a significant role. Dark matter and dark energy appear to make up a significant portion of the Universe's content. Dark matter has been introduced to explain anomalies in the motion of galaxies, while dark energy is invoked to explain the accelerating expansion of the universe.
One possibility for the existence of dark matter is domain walls \cite{Press1989,Zeldovich1974,Vilenkin1985,Avelino2008,Friedland2003}, which could have been created during the reduction of the symmetry of the initially symmetric interactions present in the early universe. Such a reduction could be supposed to be a consequence of the cooling of the initially hot universe.
It turns out if the axion field
constitutes the reason for the formation of these structures was, for example,  then it would even be possible to directly observe these objects through the effects associated with their interaction with ordinary matter as the walls pass through our galaxy \cite{Pospelov2013}. The very existence of axions was postulated to explain the strong CP problem in quantum chromodynamics, and later these particles appeared in the context of dark matter \cite{Wilczek1978,Weinberg1978,Preskill1983}. In the present work, we will focus on the study of domain wall interactions with astronomical objects.

In this context, an interesting question
that emerges concerns the study of the evolution of extended topological defects in the presence of massive bodies that are sources of gravitational fields. 
This is a theme at the interface of nonlinear
science and gravitation.
Indeed, the evolution of both vortices and domain walls in Minkowski space-time has been studied for years \cite{Forster1974, Arodz1995, Arodz1998, Dobrowolski2008, Dobrowolski2009}. In $2+1$ dimensions, the evolution of initial data corresponding to kink front was studied in article \cite{Kevrekidis2018}. In particular, for comparison with a straight front, the behavior of a deformed kink front was also investigated. The interaction of planar 
kink and anti-kink fronts in the sine-Gordon model was also considered in a similar way \cite{Carretero2022} and was also connected
to the notion of higher dimensional breather
structures. In the case of a circular front, alternate contraction and expansion are observed leading ultimately to collapse already since
the early studies of~\cite{Christiansen1979, Christiansen1981, Geicke1983}. In this context, research on breather configurations has also emerged \cite{Piette1998, Caputo2013}.

For condensed matter systems, curvature, on an equal footing with the presence of inhomogeneities, has a significant impact on the evolution of topological defects. For example, for Josephson junctions, an effective equation taking into account the effect of curvature on the evolution of a fluxon was obtained in the papers \cite{Dobrowolski2012, Gatlik2021}. It turns out that the presence of a curved region along the interface corresponds to the existence of a barrier  affecting the movement of the fluxon. In this context, the influence of the curved background on the dynamics of the extended topological soliton was studied theoretically and experimentally  in the case of the Josephson junction \cite{Gatlik2023, Gorria2004, Carapella2002, Benabdallah1996, Benadallah2000, Gulevich2007}. Prototypical 
features observed in such situations are the potential reflection or transmission of the topological defects due to the interaction with the curved region of the junction. 

In the current work, 
we extend considerations to a different
setting involving the interaction of such
nonlinear structures with a gravitational
field. 
Earlier work had considered such
a coupling in a one-dimensional setting in~\cite{oldfloyd}.
Here,
we investigate the behavior of the kink front in $2+1$ and $3+1$ curved space-time. 
This situation models how a domain wall interacts with an astronomical object like a neutron star.
We study the behavior of a kink front in the field of a gravitating object with rotational symmetry. The motivation for this study is the work \cite{Ipser1984}, in which it was found that a test particle placed in the gravitational field of a membrane is repelled from it. 
On the other hand, the paper \cite{Ipser1984} describes the collapse of a spherical domain wall. This case corresponds to the situation when outside the spherical wall we have a Schwarzschild geometry while inside we have a vacuum described by a Minkowski metric. In the present work we study the course of a similar process in the case when inside there is a spherical object (of the neutron star type). The research conducted in this work allows us to answer in the negative
the question of whether the existence of a repulsive interaction between normal matter and the domain wall is able to stop the collapse of the spherical kink front.

The article is organized as follows. In the next section (section 2), we describe the background geometry in Schwarzschild  coordinates. 
Subsequently, in section 3, we explore
the collapse of a kink front in $2+1$ dimensional space-time. A significant part of this chapter is devoted to the search for a suitable effective model that allows obtaining an accurate description of the studied phenomena in the non-perturbative regime, i.e., for arbitrary masses.
This section also contains a description of the decay of the kink front into the dominant vacuum. Then, section 4 generalizes the description to the case of a spherical domain wall in $3+1$ dimensional space-time. 
Finally, section 5 summarizes the results obtained in the article and offers
some conclusions, as well as some
directions for future study. 

\section{Scalar field model in Schwarzschild geometry}
In this paper, we describe the evolution of  kink fronts in $2+1$ and $3+1$ dimensional curved space-time with a metric having the analytical form characteristic for the Schwarzschild solution in $3+1$ dimensional space-time. 
This can be seen as a simplified representation of the interaction of the domain wall with the gravitational center in the form of a neutron star. We write the action of a scalar field on the curved background as
\begin{equation}\label{S}
S =  \int d^3 x \sqrt{|g|} \, {\cal L} = \int d^3 x \sqrt{|g|} \left[ \frac{1}{2} \, g^{\mu \nu}
\partial_{\mu} \phi \partial_{\nu} \phi - V(\phi) \right] ,
\end{equation}
where the potential has the form, characteristic of the $\phi^4$ model i.e. $V(\phi) = \frac{1}{4} (\phi^2 - 1)^2$, it is considered in the
prototypical form of a double well potential
of wide relevance to numerous applications~\cite{p4book}. This potential guarantees the existence of solutions with the form of kink front. Here $(\mu)=(0,j)$ and $(j)=(1,2,3)$ in three (in Section 4) or  $(j)=(1,2)$ in two spatial dimensions. The equation of motion for a scalar field is as follows
\begin{equation}\label{64_4}
\frac{1}{\sqrt{|g|}} \,\partial_{\mu} \left( \sqrt{|g|} \,g^{\mu \nu}
\partial_{\nu} \phi \right) + \frac{\delta V(\phi)}{\delta \phi} = 0 .
\end{equation}
It is worth noting, however, here that contrary to the work of~\cite{oldfloyd}, where we allowed the
curvature to evolve along with the nonlinear field,
here we will ``freeze'' the field to an interesting metric and we will explore the dynamics of the
nonlinear field.
An interesting situation is when the (closed) domain wall surrounds the star. This case corresponds to a prototypical scenario emulating the collapse of the domain wall towards the center of the star {(i.e., shrinking of the kink)}. 
It was shown in article \cite{Ipser1984} that a domain wall of the form of a sphere collapses. On the other hand,  there is a repulsive interaction between the wall and ordinary matter. In this situation, it is interesting to ask whether the presence of a star inside a spherical domain wall is able to stop such a collapse. The studies presented below help clarify this issue, illustrating that
the relative answer is in the negative. {Nevertheless, the simulations show that the presence of a gravitating object inside the domain wall slows down its collapse.}

For a spacetime in case of  rotational symmetry in the original Schwarzschild coordinates, the metric has the form 
\begin{equation}
\label{Sch}
  d s^2 = A(r) dt^2 - B(r) d r^2 - r^2 (d \theta^2 + \sin^2 \theta d \varphi^2)  ,
\end{equation}
where the functions $A$ and $B$ are defined by equations \eqref{A} and \eqref{B},
and $r$ is radial variable. We choose functions $A(r)$ and $B(r)$ in the form that defines the Schwarzschild exterior and the Schwarzschild constant-density interior solution
\begin{equation}
\label{A}
A(r) = \left(1 - \frac{2 M}{r} \right) \Theta(r-R) +
\frac{1}{4} \left( 3 \sqrt{1 - \frac{2 M}{R}} - \sqrt{1 - \frac{2 M r^2}{R^3}}\right)^2 \Theta(R-r) ,\\
\end{equation}
and
\begin{equation}
\label{B}
    B(r) = \frac{1}{1 - \frac{2 M}{r}} \, \Theta(r-R) + \frac{1}{1 - \frac{2 M r^2}{R^3}} \, \Theta(R - r) .
\end{equation}
where $M = \frac{4}{3} \pi R^3 \rho .$ This solution describes the metric outside and inside an spherical object with mass $M$ (first we assume $M<R/2$), radius $R$ and constant density $\rho$. Moreover, inside the star we want $A(r)>0$. This leads to the condition, $r^2 > R^2 (9 - \frac{4 R}{M})$ which is satisfied whenever $M<\frac{4}{9} \,R$. For this reason we will limit our consideration to masses smaller than this value. The first terms of the expressions for functions $A$ and $B$ describe the external Schwarzschild, while the second constant-density interior Schwarzschild solution.
The assumption of constant density leads to some unphysical features (since it replaces the equation of state of matter), but on the other hand it is commonly accepted that it approximates the interior of a neutron star quite well. In this paper, we will consider $3+1$ dimensional system and  a much simpler toy model in $2+1$ dimensions, motivated by the solution presented above.
Our intention is that the 2+1 dimensional system under consideration models the behavior of a $3+1$ dimensional system. 
In any case, we consider a fixed gravitational background, ``frozen'' to the above metric. Moreover, the $2+1$ dimensional system, due to its simplicity, makes it possible to develop methods for effective description of the field system. It allows for efficient testing of simplified descriptions due to the possibility of comparing the results obtained with the results of the field model obtained based on a relatively large lattice.

\section{Circular kink front in 2+1 
dimensions{: a toy model}}
In our considerations, we will start for
simplicity and tractability purposes
with a 2+1 dimensional toy model, the analytical form of which (in its 
radial dependence) is identical to that of the solution in 3+1 dimensions. Among other things, this model makes it possible to transparently work out how to effectively describe the dynamics of the kink with a small number of dynamic degrees of freedom. The 3+1 dimensional case will be examined in the next section.
\subsection{Field description of a domain wall with rotational symmetry.}
\begin{figure}[!ht]
    \centering
    \includegraphics[height=9cm]{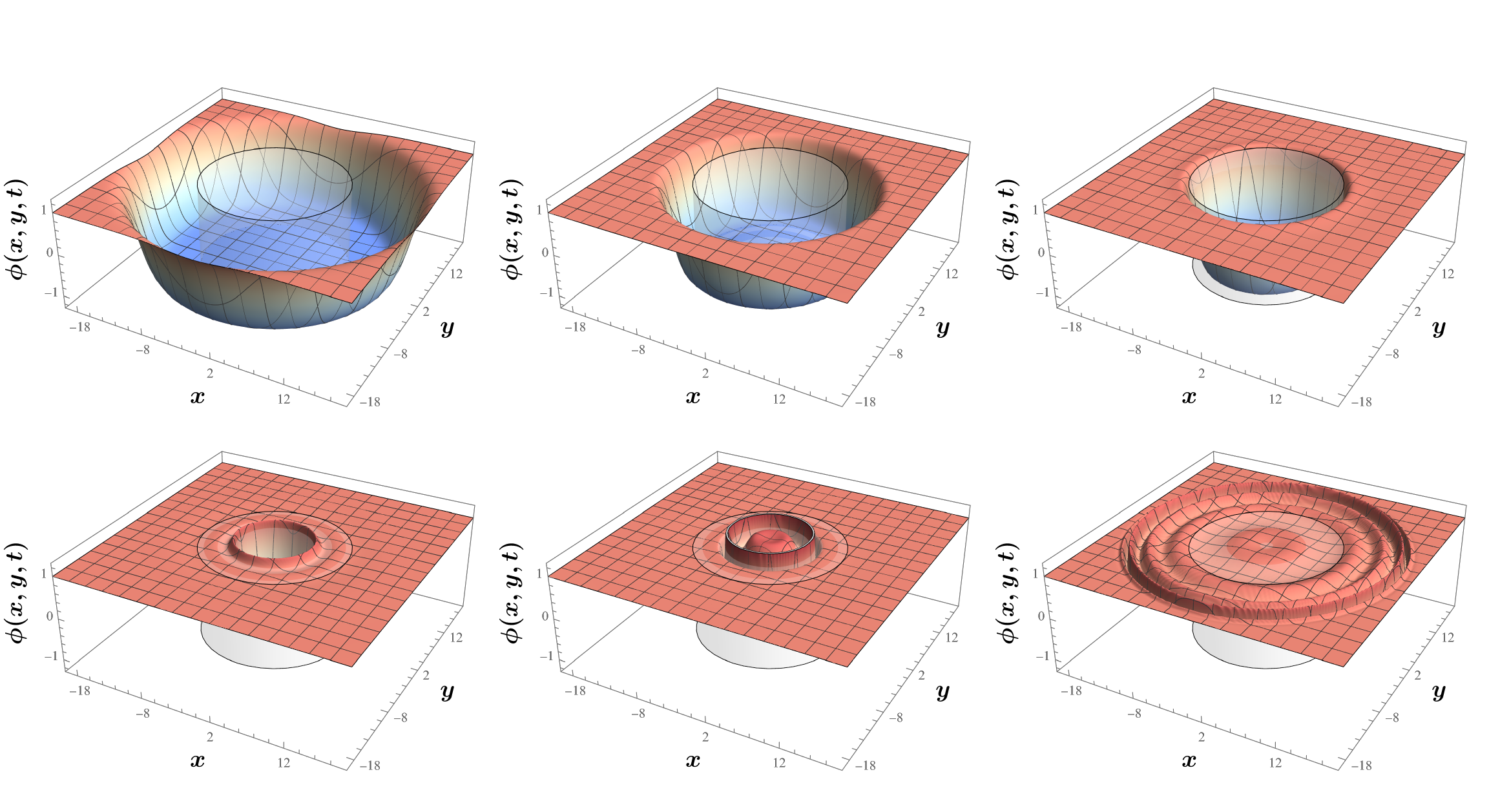}
    \caption{Kink motion in the case of $M=1$, $r_0=18$, $\dot{r}_0=0$ and $R = 10$. The gray cylinder represents a star. The $x$ and $y$ variables in the figure are Cartesian 
    coordinates in Schwarzschild geometry.
    The subsequent figures starting from the top left panel correspond to the instants $t = 0$, $15$, $20.5$, $27$, $40$ and $55$.}
    \label{fig_01}
\end{figure}

Our starting point in the 2+1 dimensional
considerations will be to adopt
Eqs.~\eqref{Sch},\eqref{A}, \eqref{B} in such a 2+1 dimensional space-time.  This 
practically reflects replacing the last term $d \theta^2 + \sin^2 \theta d \varphi^2$ in the formula \eqref{Sch} with the expression $d \varphi^2$, while leaving identical the analytical dependence of the functions $A$ and $B$ on the radial variable. This is similar to restricting considerations to the $\theta=\pi/2$ plane in $3+1$ dinensional case. In such a toy model the field equation takes the form of   
\begin{equation}
\label{radial-equation}
    \partial_t^2 \phi - \frac{1}{r} \,  G(r) \, \partial_r \left( r G(r) \partial_r \phi \right) - \frac{1}{r^2} \, A(r) \,\partial^2_{\varphi} \phi +  A(r) \phi^3 -  A(r) \phi = 0 ,
\end{equation}
where $G(r)=\sqrt{\frac{A(r)}{B(r)}}.$
Since only configurations with rotational symmetry will be studied so the term containing the derivative after the angle variable 
in numerical simulations does not play a role.
In our numerical studies we consider a domain wall moving in the gravitational field of the massive object located at the center of coordinate system. The initial configuration, inspired
by the solution of the one-dimensional
analogue of the model, is given by
\begin{equation}
\label{phi(0)}
\phi(0,r)= \tanh \left[ \frac{r-r_0}{\sqrt{2 (1-v^2)}} \right] ,
\end{equation}
\begin{equation}
\label{dphi(0)}
\partial_t \phi(0,r)=- \frac{ v}{\sqrt{2  (1 - v^2)}} \,
{\mathrm{sech}}^2 \left[ \frac{r-r_0}{\sqrt{2 (1-v^2)}} \right] .
\end{equation}
The boundary condition at the outer boundary is
$\phi(t,R_{out})=1$. This requirement is consistent with the initial condition. At the origin of the coordinate system, we assume $\partial_r \phi(t,0)=0$, which is consistent with the asymptotic value of the scalar field in the kink solution.  Such a condition is satisfied for both the kink and the dominant vacuum into which the kink could potentially collapse. The area for the simulations is the disk of radius $R_{out}=100$. In practice, when studying the evolution, we restrict ourselves to the static initial configurations of $v=0$. 
It means that, within the framework of the field model described by the equation \eqref{radial-equation}, we assume initial conditions of a circular kink with a given radius and zero initial velocity.
Figure \ref{fig_01},  shows the collapse of a domain wall with a radius larger than the radius of the star. The mass of the star is $M=1$, its radius $R=10$, while the initial radius of the wall is equal to $r_0=18$. The radius, of the star in figure \ref{fig_01} is marked with a shaded cylinder. The domain wall shrinks indefinitely towards the center of the star. 
Starting from the top left panel, Figure \ref{fig_01} shows the scalar field configurations for times $t= 0, 15$, and $20.5$, when the kink reaches the surface of the star (top row). The bottom row shows the field configurations for $t= 27, 40$ and $55$ (starting from the left). The first bottom panel (starting from left $t=27$) shows the kink approaching the center of the star while the next two show the kink decaying into the vacuum.

\subsection{Effective description of the domain wall}
Due to the symmetry, we provide an effective description of the system under study, based on the distance of the domain wall to the center of the star.
We start the construction of the effective model with an expression for the energy, which (remembering that we are considering the description in $2+1$ dimensions) has the form: 
\begin{equation}
\label{E2}
    E = \int_{\sigma_t} \sqrt{\mid h^{(2)} \mid} \, n^{\mu} \xi^{\mu} T_{\mu \nu} \,\, d^2 x ,
\end{equation}
where the energy-momentum tensor for the considered scalar field is as follows
\begin{equation}
\label{T}
    T_{\mu \nu} = \nabla_{\mu} \phi \, \nabla_{\nu} \phi - g_{\mu \nu} \, {\cal L} .
\end{equation}
The Lagrangian density ${\cal L}$ is given by formula \eqref{S} and the covariant derivative of the scalar field is the ordinary partial derivative. {Here, $\sigma_t$ is the surface of constant time. }  We denote the unit vector normal to the surface of constant time by $n^{\mu}$ and $\xi^{\nu}$ is the Killing vector corresponding to the invariance of the metric with respect to time translations. We denote the determinant of the metric induced on a constant time surface $\sigma_t$ by $h^{(2)}$. The metric tensor in $2+1$ space-time and the induced metric on the considered constant-time surface are as follows
\begin{eqnarray}
 \left[g_{\mu \nu} \right]  = \begin{bmatrix}
      A(r) & 0 & 0 \\
      0 & - B(r) & 0 \\
      0 & 0 & -r^2 
  \end{bmatrix} ,  \,\,\, \,  
   \left[h^{(2)}_{i j} \right]  = \begin{bmatrix}
      - B(r) & 0 \\
       0 & -r^2 
  \end{bmatrix} \, .
\end{eqnarray}
Here, $A(r)$ is given by the formula \eqref{A}, while $B(r)$  is given by the formula \eqref{B}.
The determinant of the induced metric on the surface of constant time is equal to $h^{(2)}=B(r)$.
In the case under consideration the normal vector and the Killing vector assume a particularly simple form $n^{\mu} = [1/\sqrt{A(r)}, 0, 0 ] $ and $\xi^{\mu} = [1, 0, 0 ] . $  After taking into account the above facts, the expression for energy reduces to 
\begin{equation}
\label{E2-r}
    E = \int_{0}^{\infty} \int_0^{2 \pi} \sqrt{\frac{B}{A}} \, T_{0 0} \,\, r d  r d \varphi .
\end{equation}
Then we use formula \eqref{T} to obtain the energy density and use  rotational symmetry of the system under consideration. After integrating over the angular variable, we obtain the following
\begin{equation}
\label{E2-r1}
    E = 2 \pi \int_{0}^{\infty} \sqrt{\frac{B}{A}} \left[ \frac{1}{2} \, (\partial_t \phi)^2 + \frac{1}{2} \frac{A}{B}\, (\partial_r \phi)^2 + A V(\phi)\right]  \,\, r d  r  ,
\end{equation}
where $V(\phi) = \frac{1}{4} (\phi^2 -1)^2$. The next step in obtaining a
reduced effective model 
 describing the dynamics of the kink
center
is to insert an ansatz of the kink form into the above expression
\begin{equation}
\label{phi_k}
    \phi_K (t,r) =  \tanh \zeta(t,r) .
\end{equation}
At this stage the only function responsible for the form of energy, and yet to be determined, is $\zeta=\zeta(t,r)$
\begin{equation}
\label{E2-r2}
    E = 2 \pi \int_{0}^{\infty} \sqrt{\frac{B}{A}} \, (\partial_{\zeta} \phi_K)^2 \left[ \frac{1}{2} \, (\partial_t \zeta)^2 + \frac{1}{2} \frac{A}{B}\, (\partial_r \zeta)^2 + \frac{1}{4} A \right]  \,\, r d  r  .
\end{equation}
 Here we used the fact that for configuration \eqref{phi_k} the potential is proportional to the square of the derivative of the $\phi_K$ function with respect to the $\zeta$ variable.
In the following sections, we argue that the proper choice of $\zeta$ is crucial to the possibility of obtaining a correct effective model that allows to describe the motion of the kink front (domain wall) in the considered system.
\subsubsection{Model based on nonrelativistic ansatz}
First, we describe the domain wall only by its position relative to the center of the star and use the ansatz 
\begin{equation}
\label{zeta_1}
     \zeta(t,r) = \frac{1}{\sqrt{2}} \left( r - r_0(t) \right) ,
\end{equation}
where $r_0(t)$ is the position of the kink. We remember that the kink front has the shape of a circle (here with radius $r_0$).
After performing integrals, we obtain an expression for the energy which resembles the energy of a point particle in an effective potential $V_1$
\begin{equation}
\label{E_eff}
    E = \frac{1}{2} \, {\cal M}_1(r_0) \, \Dot{r}_0^2 + V_1(r_0) .
\end{equation}
To save space in this section, we have included the corresponding expressions for mass ${\cal M}_1$ and potential $V_1$ in Appendix B.
The effective Lagrangian for the system under consideration 
\begin{equation}
\label{L_eff_2}
    L_{eff} = \frac{1}{2} \, {\cal M}_1(r_0) \, \Dot{r}_0^2 - V_1(r_0) ,
\end{equation}
is the basis for determining the equation of motion describing the dynamics of the collective degree of freedom
\begin{equation}
\label{19}
   {\cal M}_1(r_0) \, \Ddot{r}_0 + \frac{1}{2} \,  \frac{\partial {\cal M}_1(r_0)}{\partial r_0} \, \Dot{r}_0^2  +  \frac{\partial V_1(r_0)}{\partial r_0} = 0 .
\end{equation}
\begin{figure}[!h]
    \centering
    \subfloat{{\includegraphics[height=5cm]{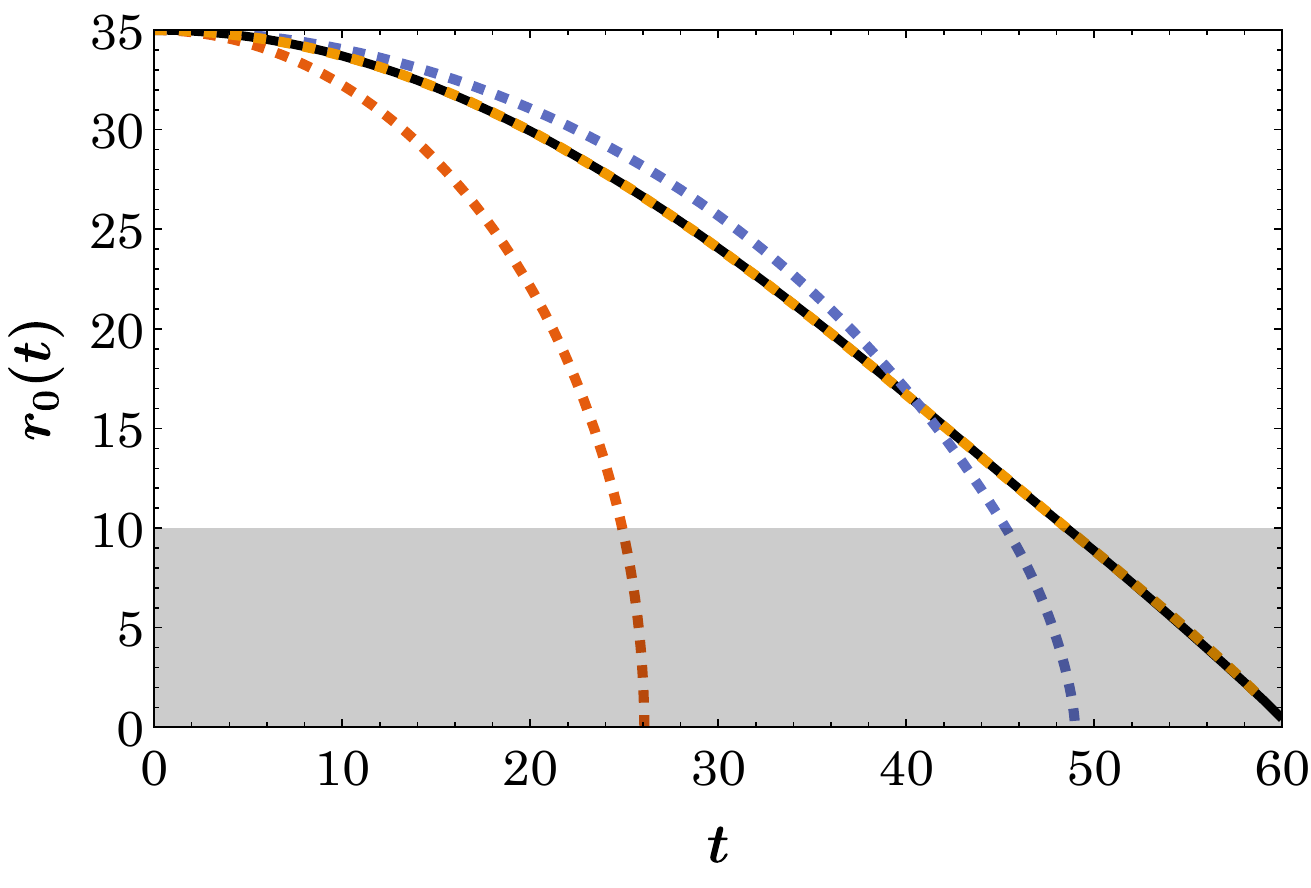}}}
    \qquad
    \subfloat{{\includegraphics[height=5cm]{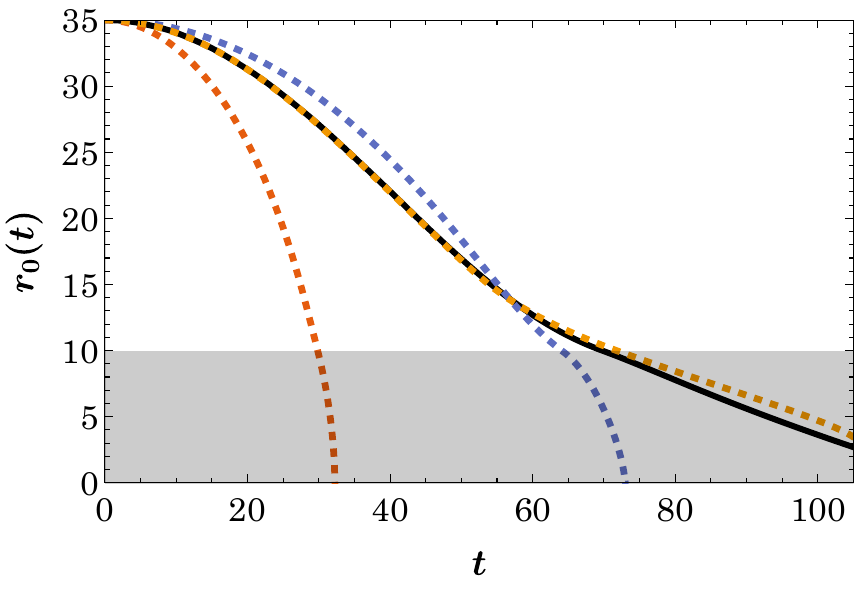}}}
    \caption{Comparison of the position of the center of the kink of the solutions of the field model (black solid line) with the first approximate model  (orange dotted line), the second model with the modified ansatz (blue dotted line) and the third model based on two degrees of freedom (yellow dotted line). In this case,  $R=10$, $r_0 = 35$ and $\Dot{r}_0 = 0$, on the left $M=1$ and on the right $M=4$.  For the third model, $\gamma(t=0) =1$. The gray area  indicates the position of the star.}
    \label{fig_02}
\end{figure}

The movement of the position of the kink front obtained on the basis of the above equation and the field model \eqref{radial-equation}
is presented in figure \ref{fig_02}. Simulations are made for the masses $M=1$ on the left  and  $M=4$ on the right. At the initial instant, the front rests at a distance $r_0 = 35$ from the gravitating center while the star radius is $R=10$. 
The trajectories of the center of the kink ($\phi=0$) obtained from the field model are represented with a black solid line while the trajectories of the kink front obtained from the effective model are represented with a dashed orange line.
It can be seen that the simple non-relativistic model shows agreement only at the very beginning of evolution. At later moments, the model only confirms the fact (and not the moment) of the collapse.
Notice that the figure also incorporates
variants of the model discussed below
offering considerably improved agreement
with the theory to which we now turn.

\subsubsection{Model based on modified ansatz}
A potentially improved ansatz should take into account the effect of gravity on the kink front.
Although, as in the previous part, we describe the domain wall only by its position relative to the center of the star, this time we use the ansatz 
\begin{equation}
\label{zeta_2}
  \zeta(t,r) = \frac{1}{\sqrt{2}} \, \sqrt{B(r_0)} \left( r - r_0(t) \right) .
\end{equation}
where $r_0(t)$ is the position of the kink. 
The inspiration for such an ansatz stems from the observation that for a flat space, the presence of a constant factor $A=const$ preceding the terms derived from the potential results in a factor $\sqrt{A}$ in the kink solution. On the other hand, if one considers the equation \eqref{radial-equation} with $\partial_{\varphi} \phi=0$ and assume a constant factor $G=const$ then it can be brought, by rescaling the radial variable, to the term appearing in flat space. Similar rescaling of the radial variable in the ansatz results in  the multiplication of the coefficient $\sqrt{A}$ by the scaling  factor $G=\sqrt{B/A}$ which, when gives $\sqrt{B}$. 
However, since in the general case the $B$ factor should take into account the effect of gravity on the kink at its present location, so we write $\sqrt{B(r_0)}$ in the ansatz. The calculations are analogous to those carried out in the previous section, and moreover, the resulting equation of motion differs only in the form of the coefficients from the previous one
\begin{equation}
\label{eff2}
   {\cal M}_2(r_0) \, \Ddot{r}_0 + \frac{1}{2} \,  \frac{\partial {\cal M}_2(r_0)}{\partial r_0} \, \Dot{r}_0^2  +  \frac{\partial V_2(r_0)}{\partial r_0} = 0 .
\end{equation}
Functions ${\cal M}_2$ and $V_2$  are also given in Appendix B. As before, the results obtained based on the above model are compared with those obtained from the field equation. In Figure \ref{fig_02}, the trajectory obtained from the field model is represented by a black solid line, while the trajectory obtained with the \eqref{eff2} model is represented by a dashed blue line. It can be seen that the course of this trajectory is much longer similar to the trajectory of the field model. Interestingly, for large masses, the effective model can be seen to bear
an imprint suggestive of a partial deflection of the trajectory when
proximal to the star.
It should be noted that we decided to modify the ansatz in this way with a factor describing the effect of gravity at the kink location because simply attaching a $\gamma(t)$ factor describing the second degree of freedom in place of $\sqrt{B(r_0)}$ did not substantially  improve the results.
\subsubsection{Model based on modified ansatz with two degrees of freedom}
\begin{figure}[!ht]
    \centering
    \subfloat{{\includegraphics[height=5.5cm]{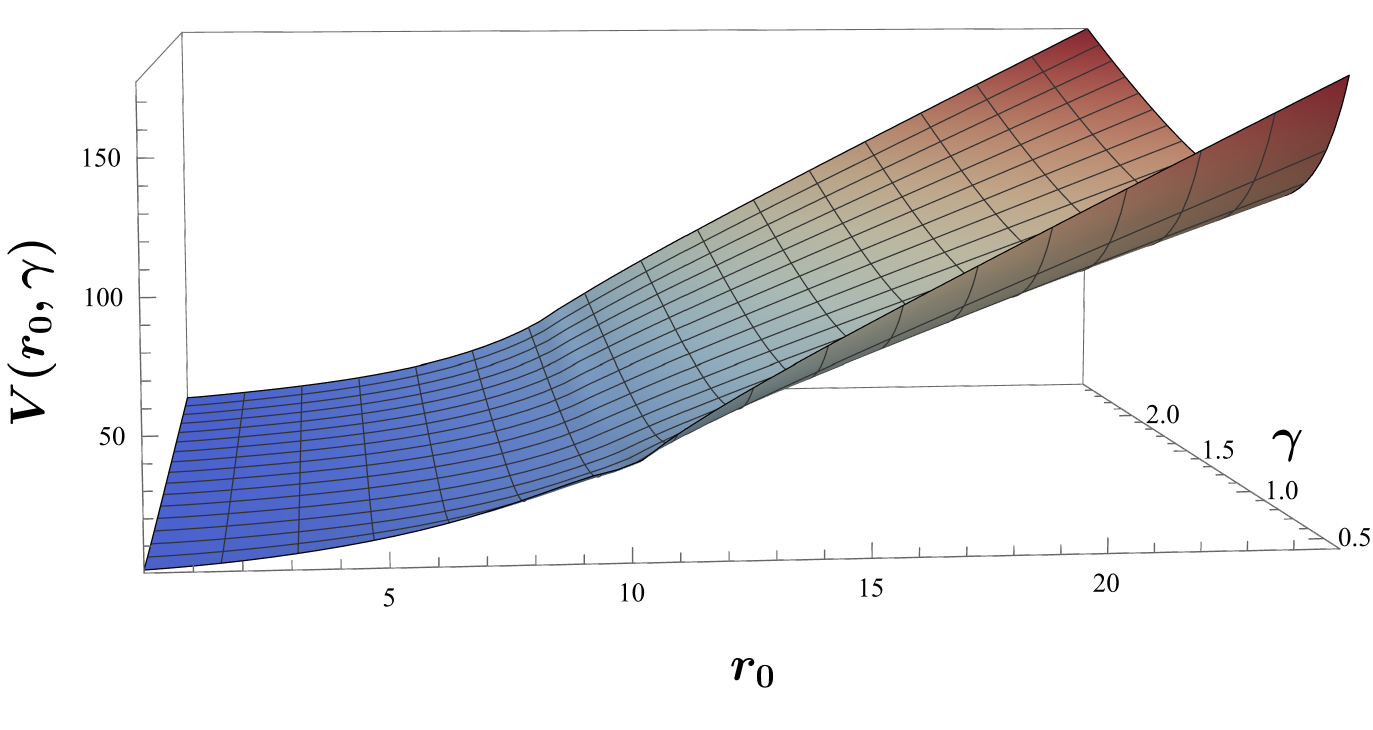}}}\\
    \qquad
    \subfloat{{\includegraphics[height=5cm]{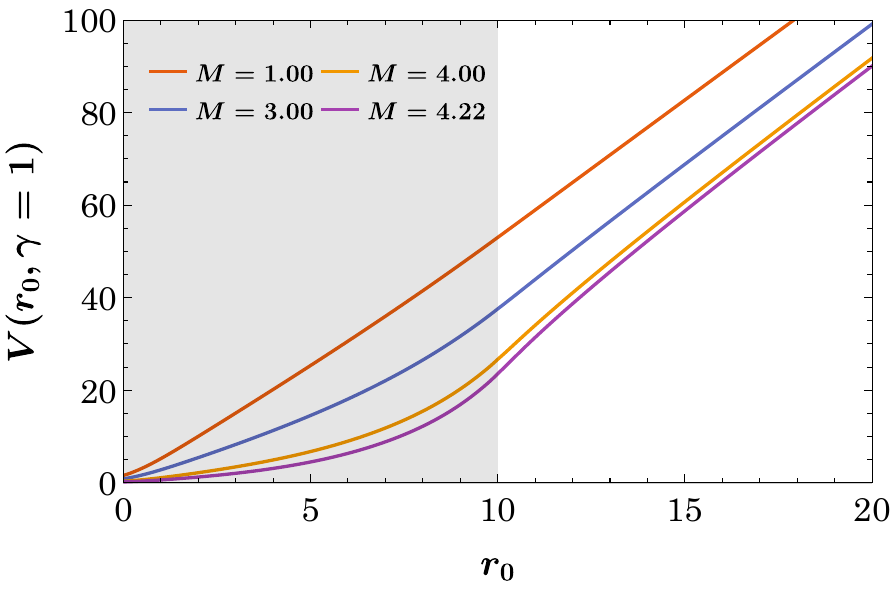}}}
    \qquad
    \subfloat{{\includegraphics[height=5cm]{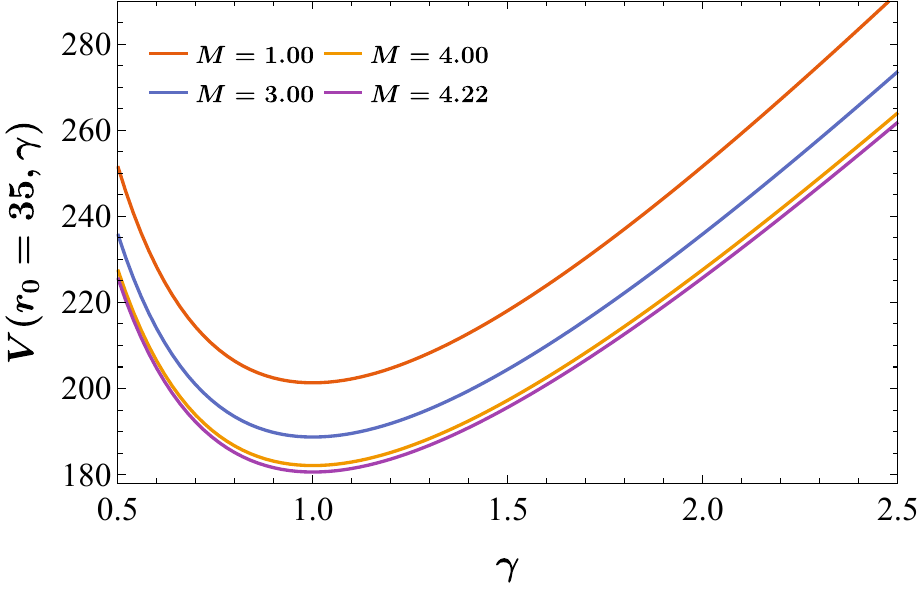}}}
    \caption{Energy landscape in the 2+1D case with rotational symmetry. The position of the star is marked as the gray area in the figure. }
    \label{fig_03}
\end{figure}

It turns out that a significant improvement in performance can be obtained if, in addition to gravitational effects, we include effects not only related to the kinematics of the front but also to the dynamics of the kink width. In order to include these effects in the ansatz we introduce an additional independent variable $\gamma=\gamma(t)$  
\begin{equation}
\label{zeta_3}
  \zeta(t,r) = \frac{1}{\sqrt{2}} \, \sqrt{B(r_0)} \, \gamma(t) \left( r - r_0(t) \right) .
\end{equation}
As in the previous two parts, we insert the ansatz into the  equation \eqref{E2-r2} obtaining 
an energy expression of the form:
\begin{equation}
\label{E_eff2}
    E = \frac{1}{2} \, {\cal M}_3(r_0, \gamma) \, \Dot{r}_0^2 + \frac{1}{2} \, a(r_0, \gamma) \, \Dot{\gamma}^2 + b(r_0, \gamma) \, \Dot{\gamma} \Dot{r}_0  +  V_3(r_0,\gamma) .
\end{equation} 
Expressions for the coefficients are provided once 
again in Appendix B.
The form of the potential $V_3=V_3(r_0,\gamma)$ is presented in Figure \ref{fig_03}. 
An important characteristic of the potential is the star mass $M$ appearing in the definition of the metric (see equations \eqref{A} and \eqref{B}). It can be seen that for masses located in the permitted area 
i.e. $M= 1.00$, $3.00$, $4.00$ and $4.22$ the only minimum of potential is located at the center of the star (the region inside the star is marked as a gray area in the  left-bottom figure). The radius of the star $R$ in the figure is $10$. The left-bottom panel of the figure shows the dependence of the potential on the radial variable, i.e. on the radius of the kink front. It can be seen that the only minimum of the potential occurs at the origin of the coordinate system, which means, a complete collapse of the kink front,
even though the decreased slope of the relevant
potential indicates that this process is slowed
(even though not stopped) by the presence of the
star. The right-bottom panel shows the dependence on the second variable. As can be seen from the figure, we have the existence of a minimum, which in turn is associated with the occurrence of oscillations of the thickness of the kink around the minimal value,
as well as the identification of a preferred
kink width (associated with the energetic
minimum). 
 The overall form of the effective potential is shown in the top panel.
Based on the expression for energy, we can provide the form of the effective Lagrangian for this system
\begin{equation}
\label{L_eff2}
    L = \frac{1}{2} \, {\cal M}_3 \, \Dot{r}_0^2 + \frac{1}{2} \, a \, \Dot{\gamma}^2 + b \, \Dot{\gamma} \Dot{r}_0 -  V_3 .
\end{equation} 
In turn, the equations of motion that describe our system take the form of
\begin{equation}
\label{eq3a}
   {\cal M}_3 \, \Ddot{r}_0 + b ~\Ddot{\gamma} +\frac{1}{2} \,  \frac{\partial {\cal M}_3}{\partial r_0} \, \Dot{r}_0^2  + \frac{\partial {\cal M}_3}{\partial \gamma} \, \Dot{\gamma} ~\Dot{r}_0  + \left( \frac{\partial b}{\partial \gamma}
   - \frac{1}{2} \frac{\partial a}{\partial r_0}\right) \Dot{\gamma}^2 + \frac{\partial V_3}{\partial r_0} = 0 ,
\end{equation}
and
\begin{equation}
\label{eq3b}
 a \, \Ddot{\gamma}  +
 b \, \Ddot{r}_0  +
 \frac{1}{2} \frac{\partial a}{\partial \gamma} \, \Dot{\gamma}^2  + 
 \frac{\partial a}{\partial r_0} \, \Dot{\gamma} \, \Dot{r}_0  + 
 \left( 
 \frac{\partial b}{\partial r_0}
   - \frac{1}{2} \frac{\partial {\cal M}_3}{\partial \gamma} 
   \right) \Dot{r}_0^2 + 
   \frac{\partial V_3}{\partial \gamma} = 0 .
\end{equation}
The trajectory in the variable $r_0$ resulting from these equations is shown in the Figure \ref{fig_02} as a yellow dashed line. The correspondence of the black line, which represents the trajectory resulting from the field equation, with the line calculated from the last model is striking. Note that the parameter $M$, which describes the deviation from flatness, is far from a small deviation to Minkowski spacetime. This result shows that the proposed ansatz goes far beyond the perturbation regime. 

\begin{figure}[!h]
    \centering
    {\includegraphics[height=8cm]{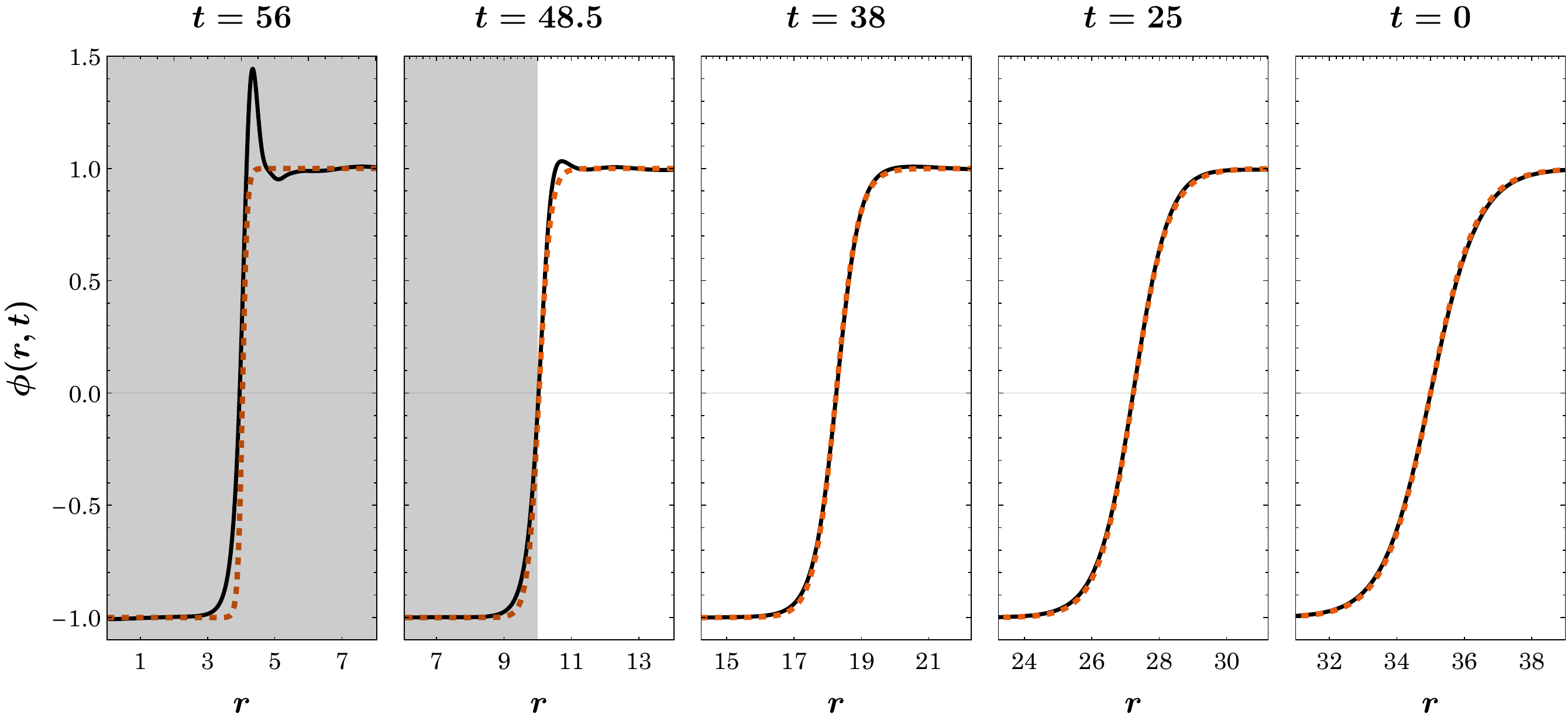}}
    \caption{Comparison of kink profiles from the field model (black line) and the third effective model based on two degrees of freedom (orange doted line). In this case, $R=10$, $r_0 = 35$,  $\Dot{r}_0 = 0$,  $\gamma(t=0) = 1$ and $M=1$. Since the kink moves towards the center (from right to left), the figures are arranged chronologically from right to left. The position of the star is marked in gray.}
    \label{fig_04}
\end{figure}

\begin{figure}[!h]
    \centering
    {\includegraphics[height=6cm]{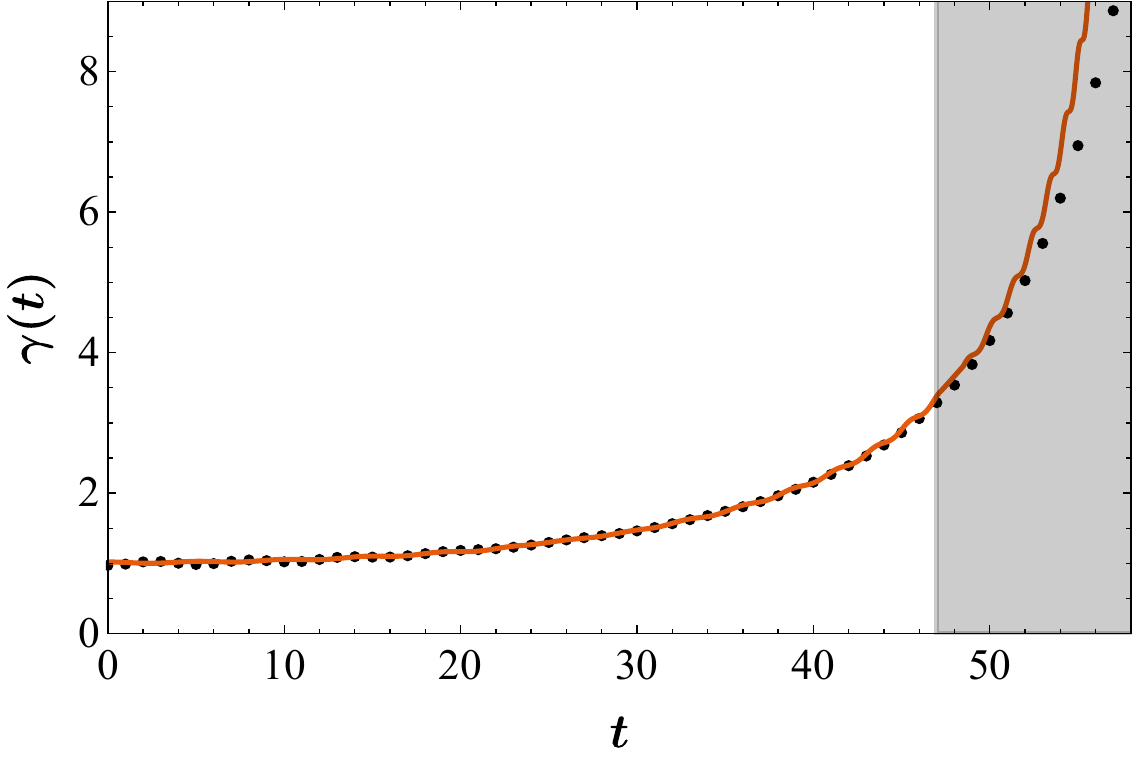}}
    \caption{Comparison of $\gamma$ value fitted from the field model (black dots) and the third effective model based on two degrees of freedom (orange doted line). In this case,  $R=10$, $r_0 = 35$,  $\Dot{r}_0 = 0$,  $\gamma(t=0) = 1$ and $M=1$. The trajectory of the kink reaching the gray field penetrates into the interior of the star.}
    \label{fig_05}
\end{figure}

We also carried out a comparison of the kink profiles, in the last model, for different moments of time with the configurations obtained on the basis of the field model. Figure \ref{fig_04} shows kink profiles at successive instants of time (red dashed line) compared with configurations obtained from the field equation (black solid line). 
Simulations are performed for a kink initially resting at $r_0=35$. 
Notice that the flow of time in the snapshots
occurse from the right (further distance of the kink
from the center at initial time) to the left
(shortest distance at the final time).
The mass of the star is $M=1$ while its radius is $R=10$. Note that the position of the zero of the scalar field in the last model and in the field model is almost identical. Moreover, there is a very high correspondence between the slopes of the kink profile in the central part (i.e. around the zero of the scalar field). Of course, towards the end of the evolution, the deformation of the field configuration is increasing which is related to the decay of the kink to the dominant vacuum described in the next section. 
Furthermore, it is important to appreciate that
the $\tanh$ ansatz utilized herein is not
capable of capturing field values $\phi>1$
arising due to the gravity-induced kink deformation
inside the star.
We would like to emphasize that despite the difference in the shape of the profiles, the positions of the zeros in both approaches are almost identical.
 Moreover, during the simulations, we also found that the presence of a star inside the domain wall has the effect of slowing down the shrinkage of the kink. This collapse is slower the more mass is inside which confirms the existence of a repulsive interaction between the domain wall and ordinary matter. The described effect is well illustrated by the animation accompanying this work \cite{animation}. 
Finally, we compared the values of the variable $\gamma$ obtained in the last effective model and the field model. The comparison is contained in Figure \ref{fig_05}. In the figure, the black points represent the results obtained from the field model while the result obtained from the effective model is described by the red line.  As in the previous figure, the kink initially rests at $r_0=35$, the mass of the gravitating object is $M=1$ while its radius is $R=10$. It can be seen that within the precision achieved (in the field model), agreement is very good. Discrepancies occur when deviations related to profile deformation are observed in the previous Figure \ref{fig_04}. As stated, these deviations appear when a kink decay takes place, i.e.  a complete change in the dynamics of the field configuration occur.

\subsection{Comment on kink decay into vacuum}
 A natural question that arises concerns the fate of the front solution as it
collapses to the origin. In this section, we show that the effect
of gravity on the decay of the kink is minor and 
practically limited
to a time shift. We also present and analyze the solution after
the collision.
\subsubsection{Kink decay in
the absence of gravity}

\begin{figure}[h!]
    \centering
    \subfloat{{\includegraphics[height=5cm]{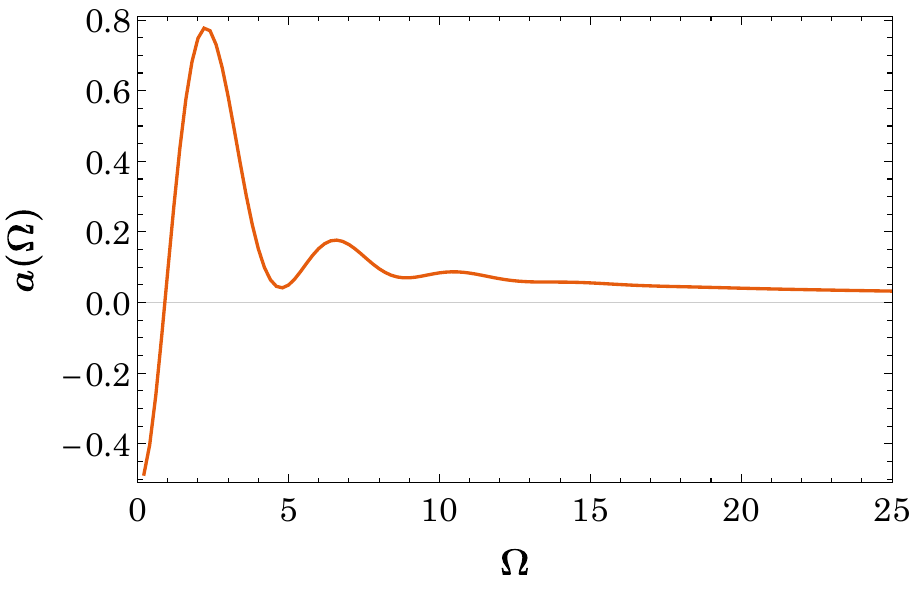}}}
    \qquad
    \subfloat{{\includegraphics[height=5cm]{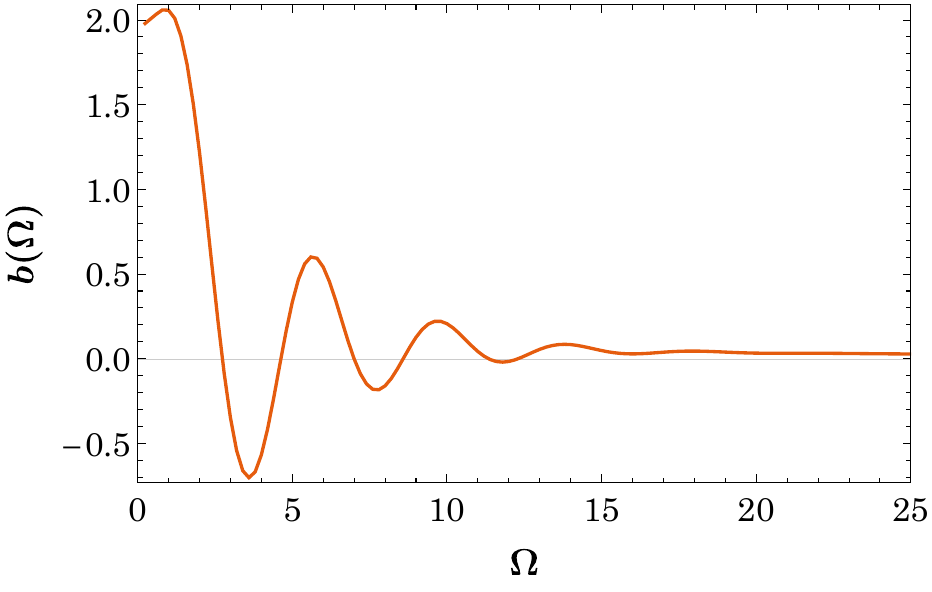}}}
    \caption{Functions $a(\Omega)$ and $b(\Omega)$ according to equations \eqref{D_09} and \eqref{D_10}.}
    \label{fig_06}
\end{figure}

In order to examine the
dynamics of the kink past the time where
it reaches the origin, we realize
that in such a setting, the dynamics
(past the reflection of the kink
at the origin) rapidly results in 
small field values around the 
homogeneous equilibrium state of
$\phi=1$. In light of this, we leverage
a linearization approach around this
equilibrium in order to offer a good
qualitative (and even semi-quantitative)
description of the relevant dynamics.
More specificially,
the equation \eqref{radial-equation} linearized around the dominant vacuum, i.e. $\phi(t,r)=1+\psi(t,r)$, in empty space ($M=0$) takes the form of
\begin{equation}
    \partial_t^2 \psi -  \frac{1}{r} \,  \, \partial_r \left( r  \partial_r \psi \right) + 2
    \psi = 0 .
    \label{D_01}
\end{equation}
The problem is linear so 
in standard way
we separate time and space variables
\begin{equation*}
\psi(t,r) = e^{i \omega t} u(r) ,
\label{D_02}
\end{equation*}
\begin{equation*}
 - \frac{1}{r} \,  \, \partial_r \left( r  \partial_r u \right) + 2   u = \omega^2 u
 .
\label{D_03}
\end{equation*}
The last equation reduces to the Helmoltz problem
with $\Omega^2 = \omega^2-2$
\begin{equation}
 - \frac{1}{r} \,  \, \partial_r \left( r  \partial_r u \right)  = \Omega^2 u .
 \label{D_04}
\end{equation}
As long as $\Omega$ is real, 
the solution is given by the Bessel function
\begin{equation*}
u(r) = J_0(\Omega r) .
\label{D_05}
\end{equation*}
Note that in the paper the condition $\omega \geq 2$ corresponds (for $M=0$) to a continuous spectrum.
An elementary solution is then
\begin{equation}
\psi(t,r) = e^{i \omega t} u(r) = e^{i \sqrt{\Omega^2+2} \, t}
J_0(\Omega r) .
\label{D_06}
\end{equation}
Using superposition, the solution to the full linearized and $M=0$
problem is
\begin{equation}
\psi(t,r) = \int_0^{\infty} J_0(\Omega r) \Omega \left[ a(\Omega)
\cos \left(\sqrt{\Omega^2 +2} \,\, t \right) +
\frac{b(\Omega)}{\sqrt{\Omega^2 +2}} \sin \left(\sqrt{\Omega^2 +2}
\,\, t \right)\right] d \Omega .
\label{D_07}
\end{equation}
The coefficients $a$, $b$ are obtained by projecting onto $J_0(\Omega r)$ in the sense of
the Sturm-Liouville inner product
\begin{equation*}
\langle f, g \rangle = \int_0^{\infty} f(r) g(r) r  dr .
\label{D_08}
\end{equation*}
As a result of the projection, we get
\begin{equation}
a(\Omega) = \int_0^{\infty} \psi(t=0,r) J_0(\Omega r) r dr ,
\label{D_09}
\end{equation}
\begin{equation}
b(\Omega) = \int_0^{\infty} \partial_t \psi(t=0,r) J_0(\Omega r) r
dr .
\label{D_10}
\end{equation}
The coefficients obtained in the above manner as functions of $\Omega$ are shown in Figure \ref{fig_06}. It can be seen that the significant dependence on $\Omega$ occurs mainly for small values of this parameter. As $\Omega$ increases, both coefficients tend to zero.
\begin{figure}[h!]
    \centering
    \subfloat{{\includegraphics[height=5cm]{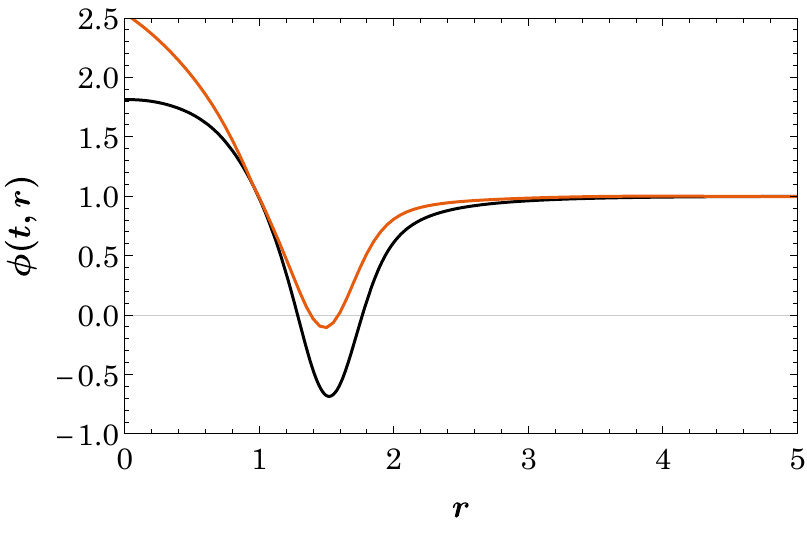}}}
    \qquad
    \subfloat{{\includegraphics[height=5cm]{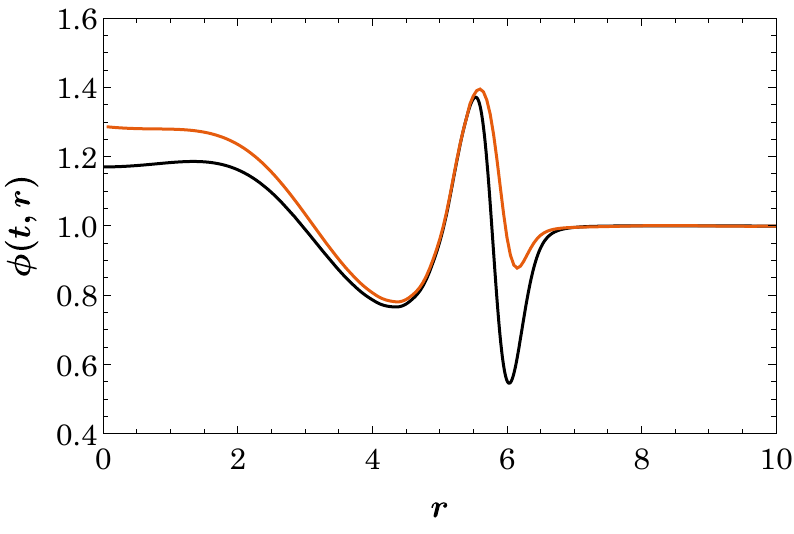}}}
    \qquad
    \subfloat{{\includegraphics[height=5cm]{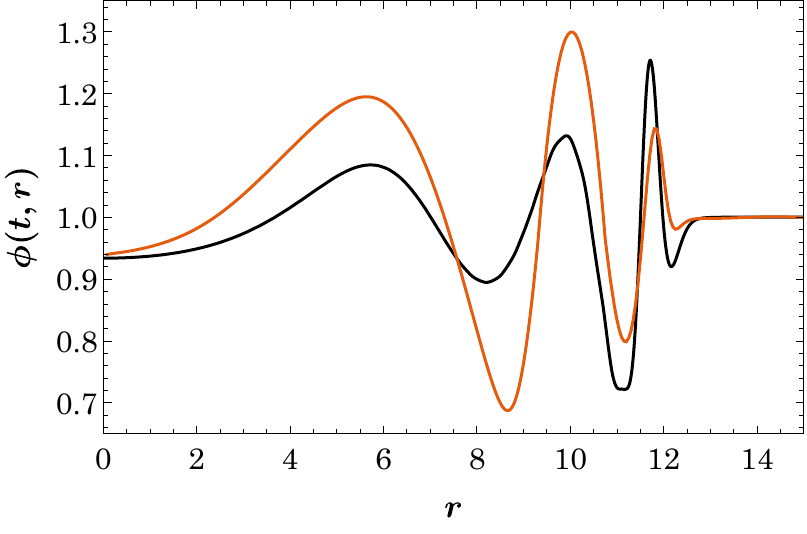}}}
    \qquad
    \subfloat{{\includegraphics[height=5cm]{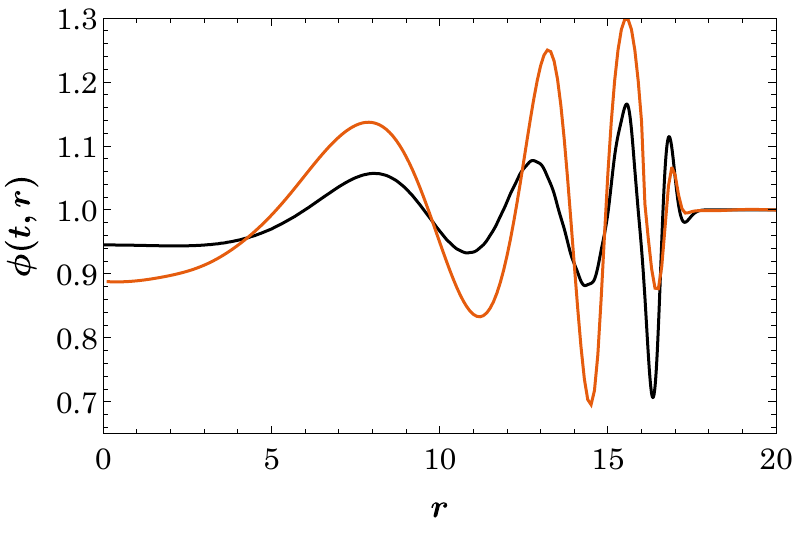}}}
    \caption{Comparison of the solutions of the full field model (black line) with the configuration base on the solutions of equation \eqref{D_07} (orange line) for times $t =0$, $4.5$, $10.5$ and $15.5$. }
    \label{fig_07}
\end{figure}
Figure \ref{fig_07} shows the decay of the kink to the dominant vacuum. The evolution of the scalar field obtained from the nonlinear field equation \eqref{radial-equation} is shown by the black line. On the other hand, the analytical result  based on the  formula \eqref{D_07} achieved for the  linear equation \eqref{D_01}  is represented by the red line. 
It should be clarified here that the black line shows the configuration obtained in the field model \eqref{radial-equation}, with the initial condition having a form of a static kink starting to evolve at a distance $r=18$ from the origin. The kink only begins to decay into the dominant vacuum when it reaches the origin of the coordinate system. This configuration is compared with the results of the linear model \eqref{D_01} describing vacuum excitations. Therefore, the times in the figure refer precisely to the model \eqref{D_01}, which means that, for example, the moment $t=0$ in the figure corresponds to t=30.1 in the field model \eqref{radial-equation}. 
The subsequent figures starting with (a) and ending with (d) show the configuration of the scalar field at times $t = 0, 4.5, 10.5$ and $15.5$ after kink decay. Note that the positions of the extremes for both curves are similar in all figures. Although discrepancies occur in the case of amplitudes, it is relevant to recall that in the linear model the amplitudes are subject to a scaling factor.
In the above sense, the compatibility of the black and red curves is really considerable, suggesting the nearly linear nature
of the corresponding dynamics. This means that indeed, at least in empty space, the linear model \eqref{D_01} confirms the decay of the kink into the vacuum state. 

\subsubsection{Kink decay in the presence of gravity.} 
\begin{figure}[h!]
    \centering
    \subfloat{{\includegraphics[height=5cm]{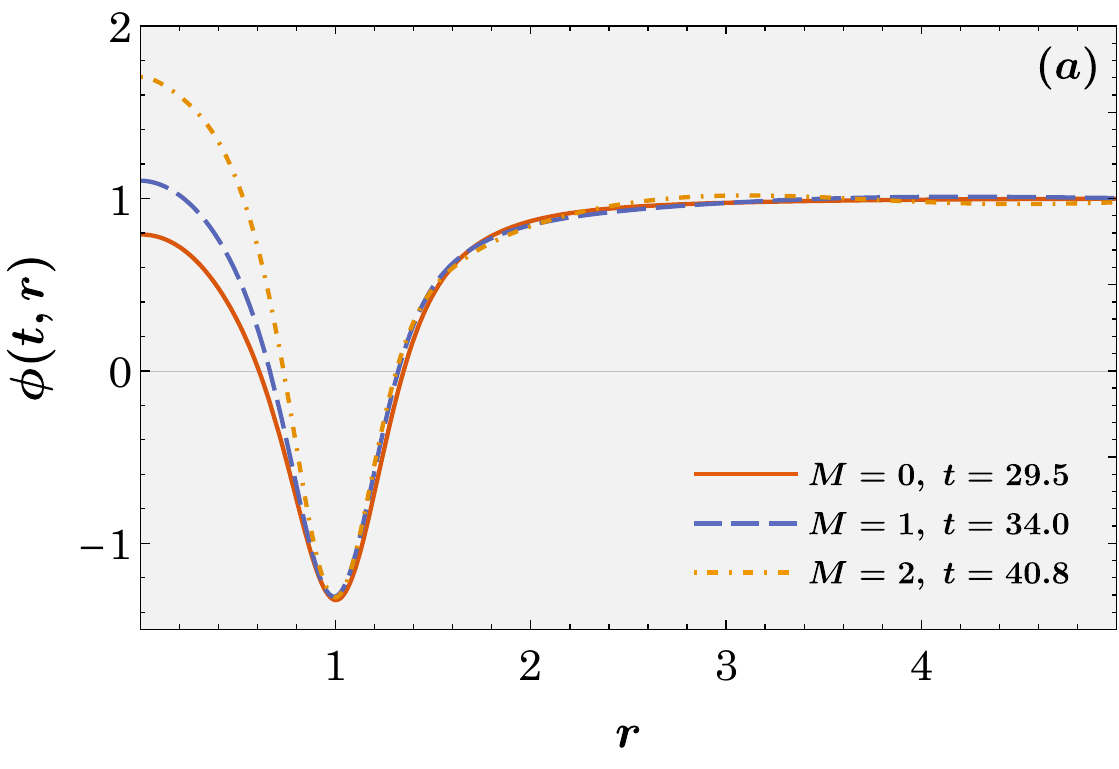}}}
    \qquad
    \subfloat{{\includegraphics[height=5cm]{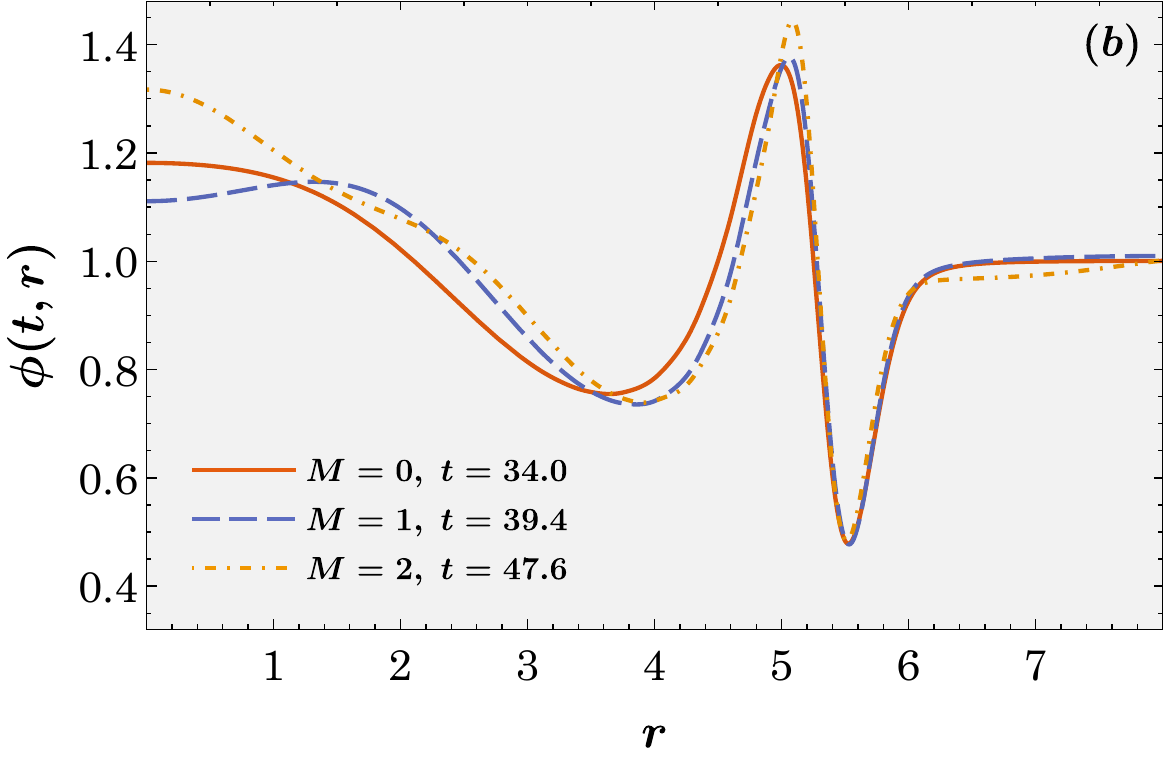}}}
    \qquad
    \subfloat{{\includegraphics[height=5cm]{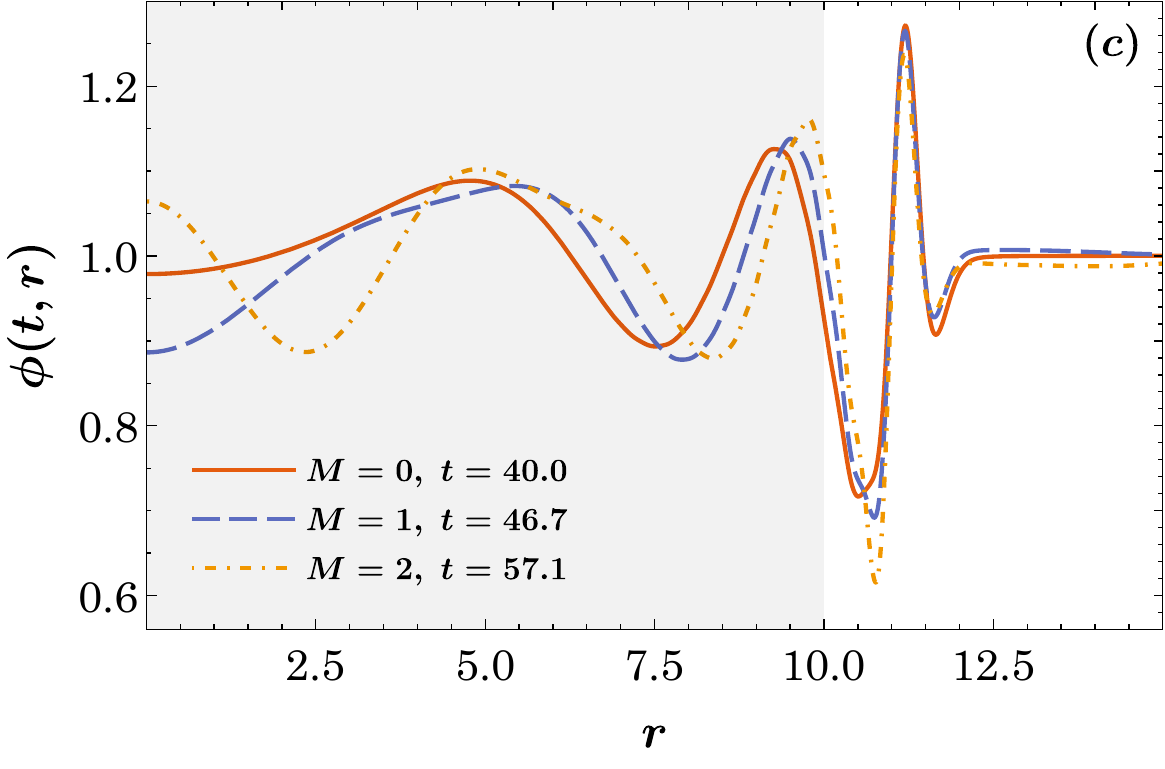}}}
    \qquad
    \subfloat{{\includegraphics[height=5cm]{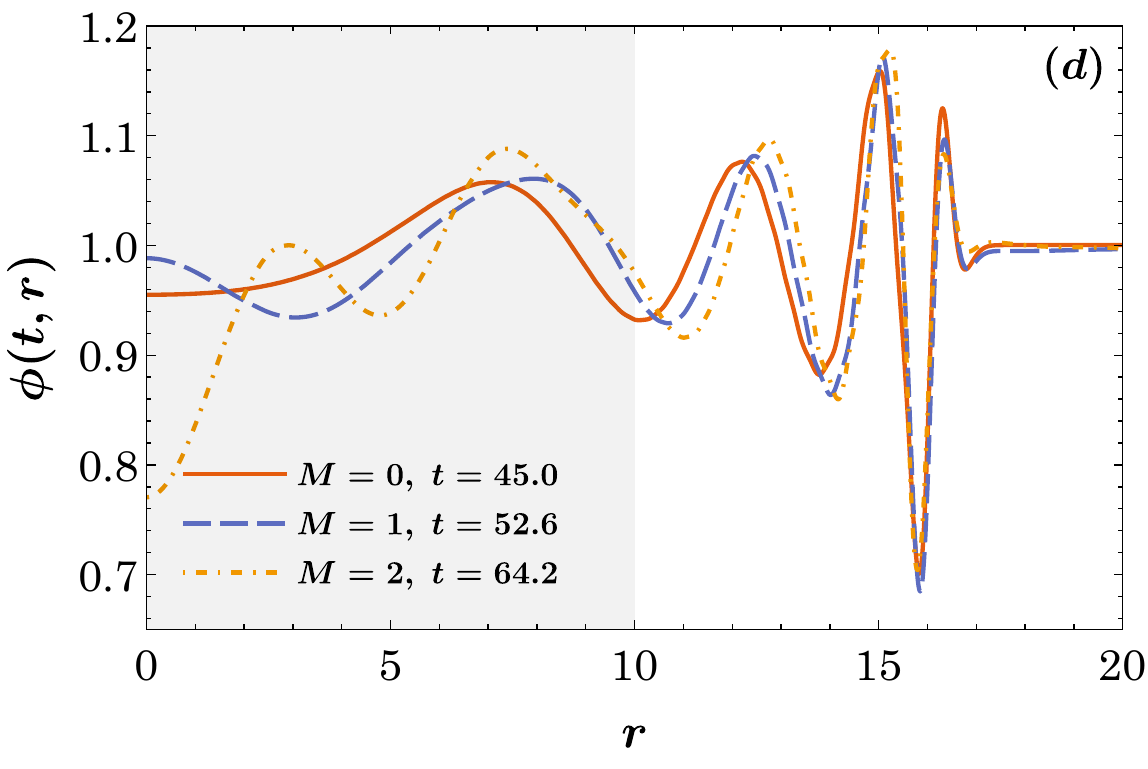}}}
    \caption{ Kink decay into vacuum for a model with no gravity ($M=0$), and two masses ($M=1$ and $M=2$), in each case $R =10$. The graphics show the situation immediately after the collision at the center of the star (a), inside the star (b), on the surface (c) and outside the star (d). In each case the star is highlighted in grey. In all simulations (for $M=0$, $M=1$ and $M=2$), at the beginning of evolution the kink rests at a distance $r=18$ from the center of the coordinate system.}
    \label{fig_08}
\end{figure}

The course of the kink decay into the vacuum is illustrated in Figure \ref{fig_01}. This figure presents the
initial shrinking of the kink (upper three panels and lower first from the left) and the subsequent outward movement of excitations from the
center (lower middle and right panels).
On the other hand, Figure \ref{fig_08} provides
a detailed representation, depicting four stages of the kink’s decay into the vacuum: (a) just after reflection from
the star’s center, (b) during decay within the star, (c) at the star’s surface, and (d) away from the star. In all the figures,
the course of evolution in the absence of gravitational interactions with the star is depicted with a  continuous red line. The other
 lines show the field configurations for the mass of a gravitating object equal to M = 1 (blue dashed line) and M = 2
(orange dashed-doted line). 
The gray area in the figures represents the interior of the star. It can be noted that the decay
process and resulting excitations are similar in amplitude and shape across all stages of evolution, differing primarily in their
timing and the precise amplitude of the tail (far from the
reflected excitation ``center'', or, e.g., to be more precise,
its point of
minimum field value).  Let us underline that each case starts with zero initial velocity (at the same position), but a time shift associated with increasing mass is observed. This shift is related to the gravitational force, dependent on the star's mass, acting on the kink before it reaches the center. 
It is relevant to point out here that the vacuum of $\phi=1$
remains a stable configuration (as we will show below),
although the spectrum of excitations of the vacuum solution in the presence of a gravitational field is definitely richer than in empty space.

\subsubsection{Linear stability of dominant vacuum}
\begin{figure}
    \centering
    \subfloat{{\includegraphics[height=5cm]{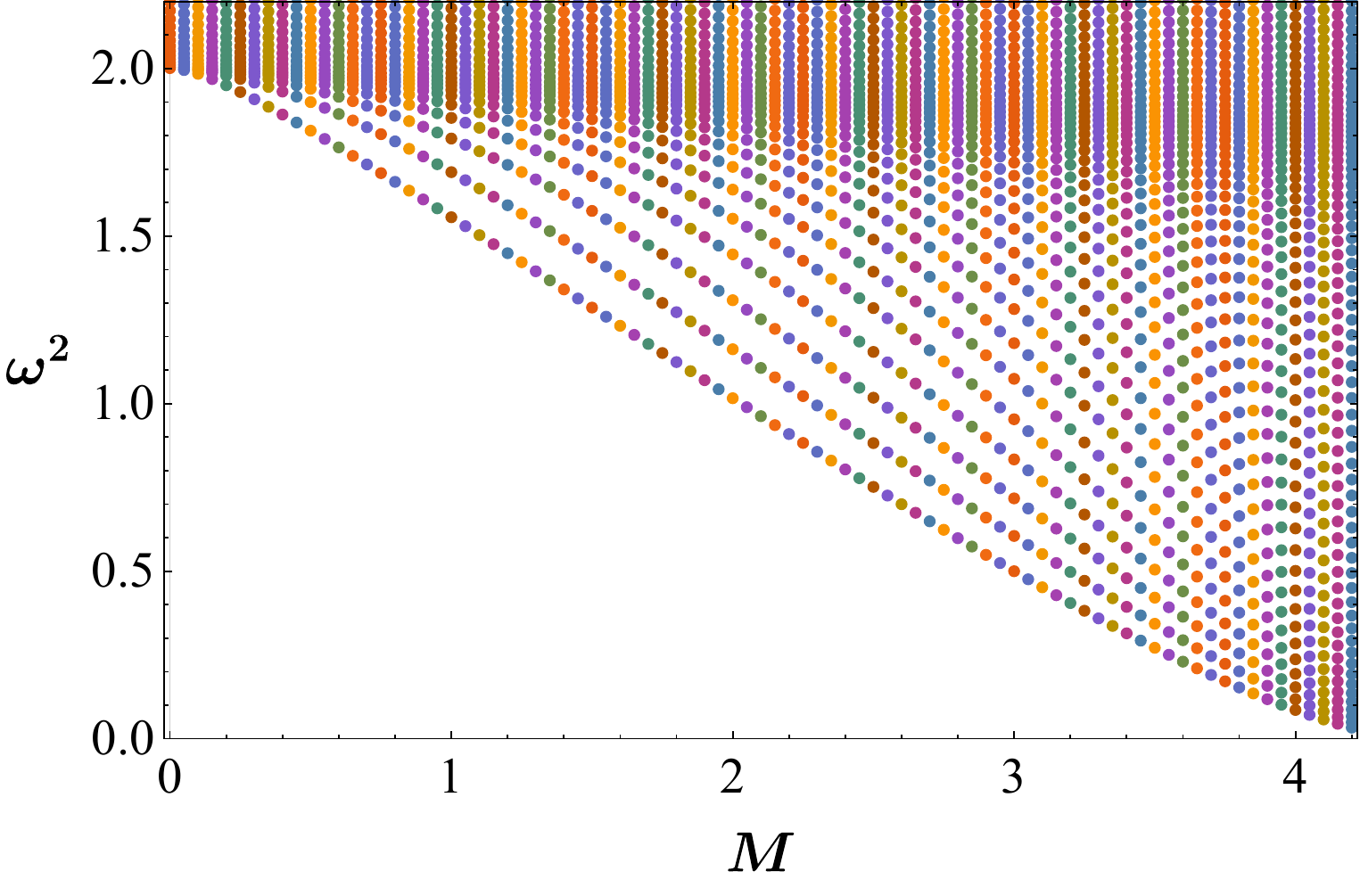}}}
    \qquad
    \subfloat{{\includegraphics[height=5 cm]{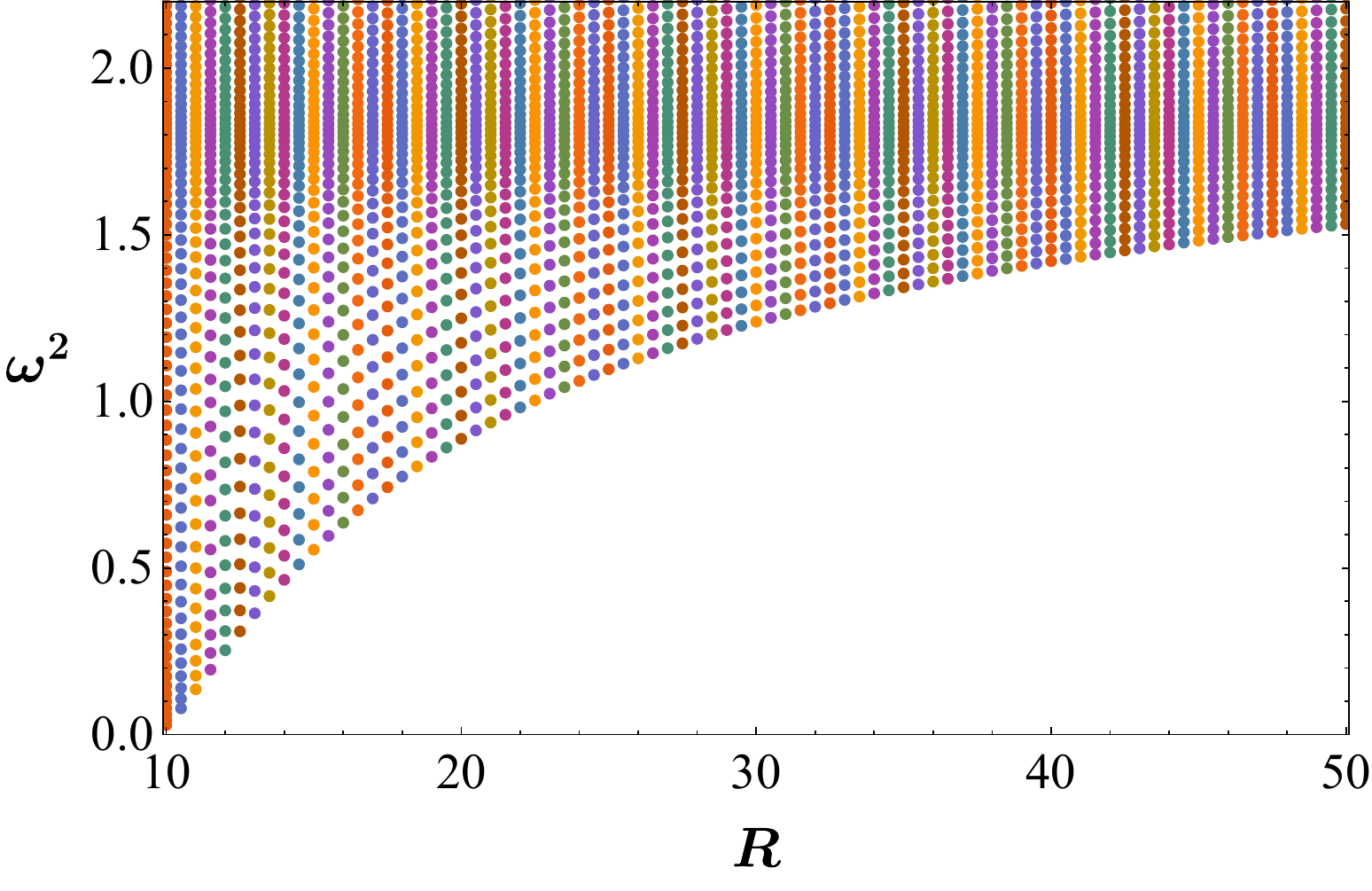}}}
    \caption{Vacuum state excitations in the $\phi^4$ model in the geometry determined by the star's gravitational field. On the left, the squared eigenfrequencies are shown as a function of the mass of the star for a fixed  radius $R=10$. On the right panel, the squared eigenfrequencies are shown as a function of the radius of the star $R$ for a mass  $M = 4.22$.}
    \label{fig_09}
\end{figure}

As discussed in the previous subsection, the dynamics results
in the decay of the kink structure to
$\phi=1$, i.e., to the vacuum value reached by the scalar field at spatial infinity. The stability of this vacuum in the considered geometry can be studied by examining the linearized equation  $\phi(t,r)=1 + \psi(t,r)$. Inserting this expansion into the equation of motion \eqref{radial-equation} yields, with the accuracy to the linear terms in the perturbation
\begin{equation}
\label{psi}
\partial_t^2 \psi - \frac{1}{r} \, G \, \partial_r \left( r G \partial_r \psi \right) + 2  A \psi = 0 .
\end{equation} 
Due to the radial symmetry of the system and the fact that only radial perturbations are being studied, the term dependent on the angular variable was zeroed out; the latter term, through a 
Fourier mode decomposition, would
result in a shift in the frequencies considered in what follows. By confining ourselves to perturbations of the form $\psi(t,r) = e^{i \omega t} u(r)$, we obtain an equation of the Schr{\"o}dinger type which describes the possible excitations of the configuration under consideration 
\begin{equation}
\label{psi-u}
 - \frac{1}{r} \, G \, \partial_r \left( r G \partial_r u \right) + 2  A u = \omega^2 u .
\end{equation}
Note that for $M=0$ the function $G=1$ ($A=1$), and thus we obtain a reduction to the vacuum stability problem in free space. In this case, the continuous spectrum appears above a threshold defined by the value of $\omega^2=2$. Note that in this case equation (28) reduces to the Bessel equation. {This case was investigated in detail in previous section}. 
On the other hand, if the mass of the star $M$ is greater than zero then we observe the appearance of stable discrete modes. 
The spectrum resulting from the above equation is shown in Figure \ref{fig_09}. The left panel shows the excitations as a function of  the mass of the star $M$ for a fixed radius of the star R=10. The right panel of the figure shows the spectrum as a function of radius $R$ for a mass $M=4.22$.  Note that the beginning of the graph located on the right panel corresponds to the end point of the drawing located on the left panel. 
We can capture the most essential features of the solutions of the above equation by taking $A$ and $G$ in the form describing the space-time outside the gravitating object. In this case, equation \eqref{psi-u} is reduced to the form
\begin{equation}
\label{psi-u2}
 - \frac{1}{r} \, \left( 1 - \frac{2 M}{r}\right) \, \partial_r \left( r \left( 1 - \frac{2 M}{r}\right) \partial_r u \right) + 2  \left( 1 - \frac{2 M}{r}\right) u = \omega^2 u .
\end{equation}
This equation particularly simplifies at spatial infinity 
($r \rightarrow \infty$)
\begin{equation}
\label{psi-as}
\partial_r^2 u + (\omega^2 - 2 ) u = 0 .
\end{equation}
Depending on the parameter $\omega$, this equation has two types of solutions. The first for $\omega^2>2$ describes the excitations belonging to the continuous spectrum.
They have the form 
\begin{equation}
\label{uas}
 u(r) = A e^{+ i \Omega r } + B e^{- i \Omega r } ,
\end{equation}
where $\Omega = \sqrt{\omega^2 - 2 }$. The second type of solutions occurs for $\omega^2<2$ and has the form
\begin{equation}
\label{uas2}
 u_n(r) = C e^{- \Delta r } 
\end{equation}
here $\Delta = \sqrt{2 - \omega_n^2}$. This function describes the asymptotic form of the solutions to the equations \eqref{psi-u2} and \eqref{psi-u}.
 Although the equation \eqref{psi-as} allows us to establish the behavior of the solutions of the \eqref{psi-u} equation at spatial infinity, the existence of bound states has its origin in the existence of a gravitational object located around zero, leading
 to localized (point spectrum) modes. 
The spectrum obtained numerically (Fig. \ref{fig_09}) shows the absence of unstable modes in the excitation spectrum of the dominant vacuum and clearly suggests that the kink which decays in 
such a radial problem, in fact, transforms into (i.e.,
disperses into) a vacuum solution.

\section{Spherical domain wall in 3+1 dimensions}
The considerations of Section 3 can easily be generalized to a system with spherical symmetry in 3+1 dimensional spacetime.
\subsection{Field description of a domain wall with spherical symmetry.}
Assuming spherical symmetry, we can write the field equation 
for the $3+1$-dimensional case in the metric \eqref{Sch}
\begin{equation}
    \partial_t^2 \phi - \frac{1}{r^2} \,  G(r) \, \partial_r \left( r^2 G(r) \, \partial_r \phi \right)  
    - \frac{1}{r^2 \sin \theta} \,  A(r) \, \partial_{\theta} \Big( \sin \theta \, \partial_{\theta} \phi \Big) 
    - \frac{1}{r^2 \sin^2 \theta} \,  A(r) \, \partial^2_{\varphi}  \phi  
    +  A(r) \phi^3 -  A(r) \phi = 0 .
\end{equation}
Due to the spherical symmetry of the mass distribution and the initial and boundary conditions, the derivatives 
with respect to the angles play no role.
We assume both initial \eqref{phi(0)}, \eqref{dphi(0)} and boundary conditions analogous to the previous section.
\subsection{Effective description of the spherical domain wall}
\begin{figure}[!ht]
    \centering
    \subfloat{{\includegraphics[height=5cm]{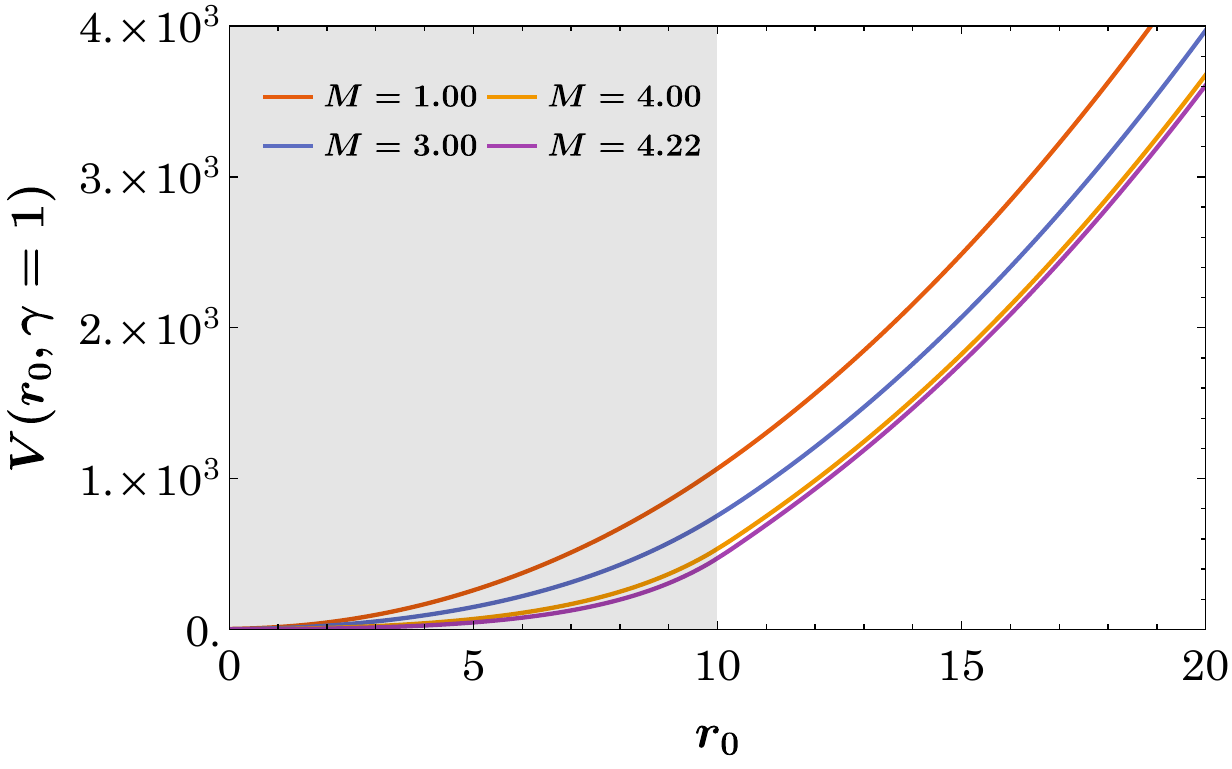}}}
    \qquad
    \subfloat{{\includegraphics[height=5cm]{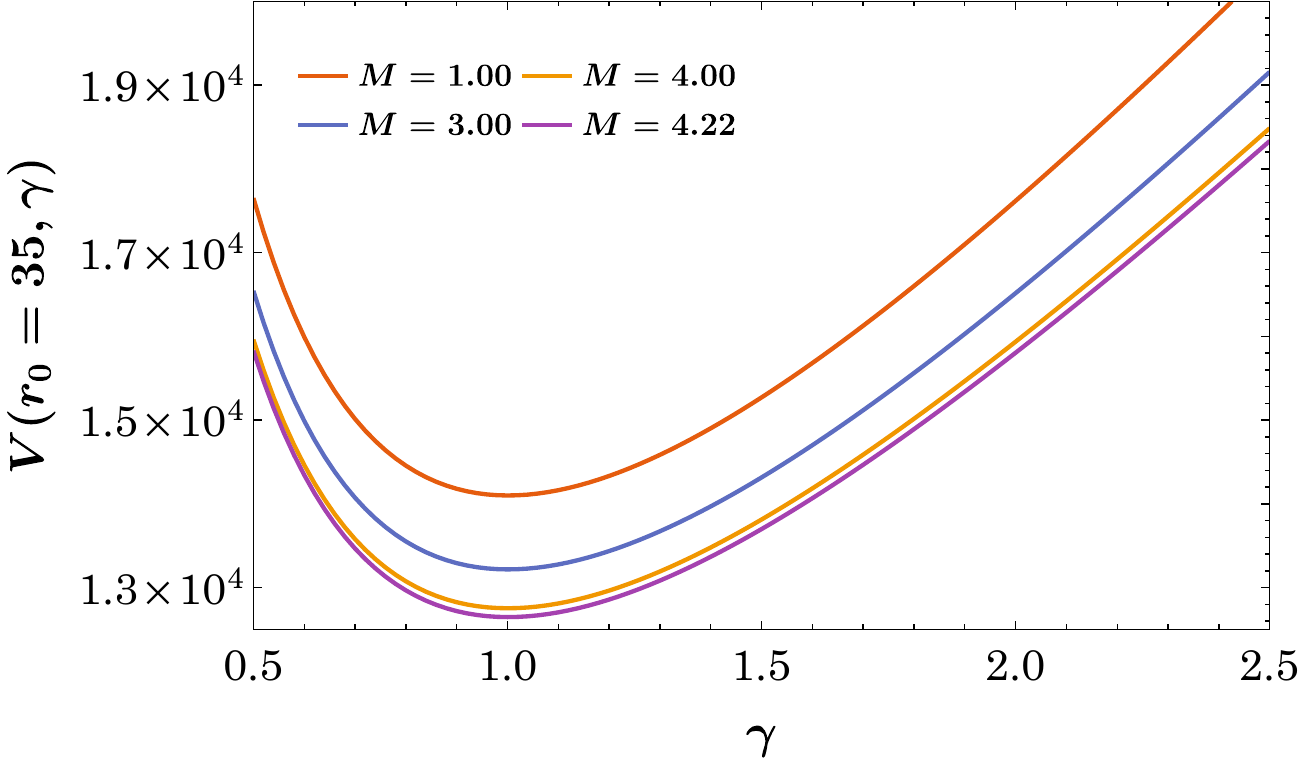}}}
    \caption{Energy landscape in the spherically symmetric 3+1D case. The grey area, in the left panel, represents the star.}
    \label{fig_10}
\end{figure}

Due to the particular simplicity of the reasoning presented in the previous section, we can easily repeat it in the 3+1 dimensional case. The energy integral in this case has the form
\begin{equation}
    E = \int_{\Sigma_t} \sqrt{\mid h^{(3)} \mid} \, n^{\mu} \xi^{\mu} T_{\mu \nu} \,\, d^3 x ,
\end{equation}
where $\Sigma_t$ denotes the three-dimensional hypersurface of constant time and $h^{(3)}$ is the determinant of the metric induced on this hypersurface. The metric this time is the full Schwarzschild metric, and the metric induced on the hypersurface $\Sigma_t$ is its spatial part
\begin{eqnarray}
 \left[g_{\mu \nu} \right]  = \begin{bmatrix}
      A(r) & 0 & 0 & 0\\
      0 & - B(r) & 0 & 0\\
      0 & 0 & -r^2 & 0 \\
       0 & 0 & 0 & -r^2 \sin^2 \theta
  \end{bmatrix} ,  \,\,\, \,  
   \left[h^{(3)}_{i j} \right]  = \begin{bmatrix}
      - B(r) & 0 & 0\\
       0 & -r^2 & 0 \\
       0 & 0 & -r^2 \sin^2 \theta 
  \end{bmatrix}  .
\end{eqnarray}
Here $(\mu)=(0,j)$ and $(j)=(1,2,3)$. Due to the form of the normal vector $n^{\mu} = [1/\sqrt{A(r)}, 0, 0, 0 ] $ and the Killing vector $\xi^{\mu} = [1, 0, 0, 0 ] $ , the energy can be written as follows
\begin{equation}
    E = \int_{0}^{\infty} \int_0^{2 \pi} \int_0^{\pi} \sqrt{\frac{B}{A}} \, T_{0 0} \,\, r^2 \sin \theta d  r d \varphi d \theta \, .
\end{equation}
Due to rotational symmetry, after inserting the expression for energy density, we get
\begin{equation}
\label{E2-r3}
    E = 4 \pi \int_{0}^{\infty} \sqrt{\frac{B}{A}} \left[ \frac{1}{2} \, (\partial_t \phi)^2 + \frac{1}{2} \frac{A}{B}\, (\partial_r \phi)^2 + A V(\phi)\right]  \,\, r^2 d  r  ,
\end{equation}
where we have performed the integration with respect to the angular variables. 
As before we insert an ansatz of the kink form \eqref{phi_k} into the above expression obtaining
\begin{equation}
\label{E2-r4}
    E = 4 \pi \int_{0}^{\infty} \sqrt{\frac{B}{A}} \, (\partial_{\zeta} \phi_K)^2 \left[ \frac{1}{2} \, (\partial_t \zeta)^2 + \frac{1}{2} \frac{A}{B}\, (\partial_r \zeta)^2 + \frac{1}{4} A \right]  \,\, r^2 d  r  .
\end{equation}
Then we adopt the form of the function $\zeta$ given by the formula \eqref{zeta_3}. As a result of this operation, we get an expression for the energy that is analogous to the expression \eqref{E_eff2}
\begin{equation}
\label{E_eff33}
    E = \frac{1}{2} \, {\cal M}(r_0, \gamma) \, \Dot{r}_0^2 + \frac{1}{2} \, \Tilde{a}(r_0, \gamma) \, \Dot{\gamma}^2 + \Tilde{b}(r_0, \gamma) \, \Dot{\gamma} \Dot{r}_0  +  V(r_0,\gamma) .
\end{equation} 
The coefficients appearing in this expression are defined in Appendix C. The potential appearing in the above expression has all the same features as the potential obtained with the same ansatz in the $2+1$ dimensional case. The potential is shown in Figure \ref{fig_10}. The left panel shows the dependence on the radial variable. It can be seen that the only minimum here occurs at the origin of the coordinate system. This indicates an 
attraction of the kink towards the origin
(that cannot be avoided ---but only delayed--- due to the
presence of the star), and consequently a decay of the domain wall into the dominant vacuum. The right panel, on the other hand, shows the dependence of the potential on the $\gamma$ variable. The minimum that occurs in this figure shows the tendency for the width of the domain wall to oscillate around a certain value. Putting the two figures together gives rise to an effective potential inclined toward zero radius and bearing a finite width.

\begin{figure}
    \centering
    \subfloat{{\includegraphics[height=5cm]{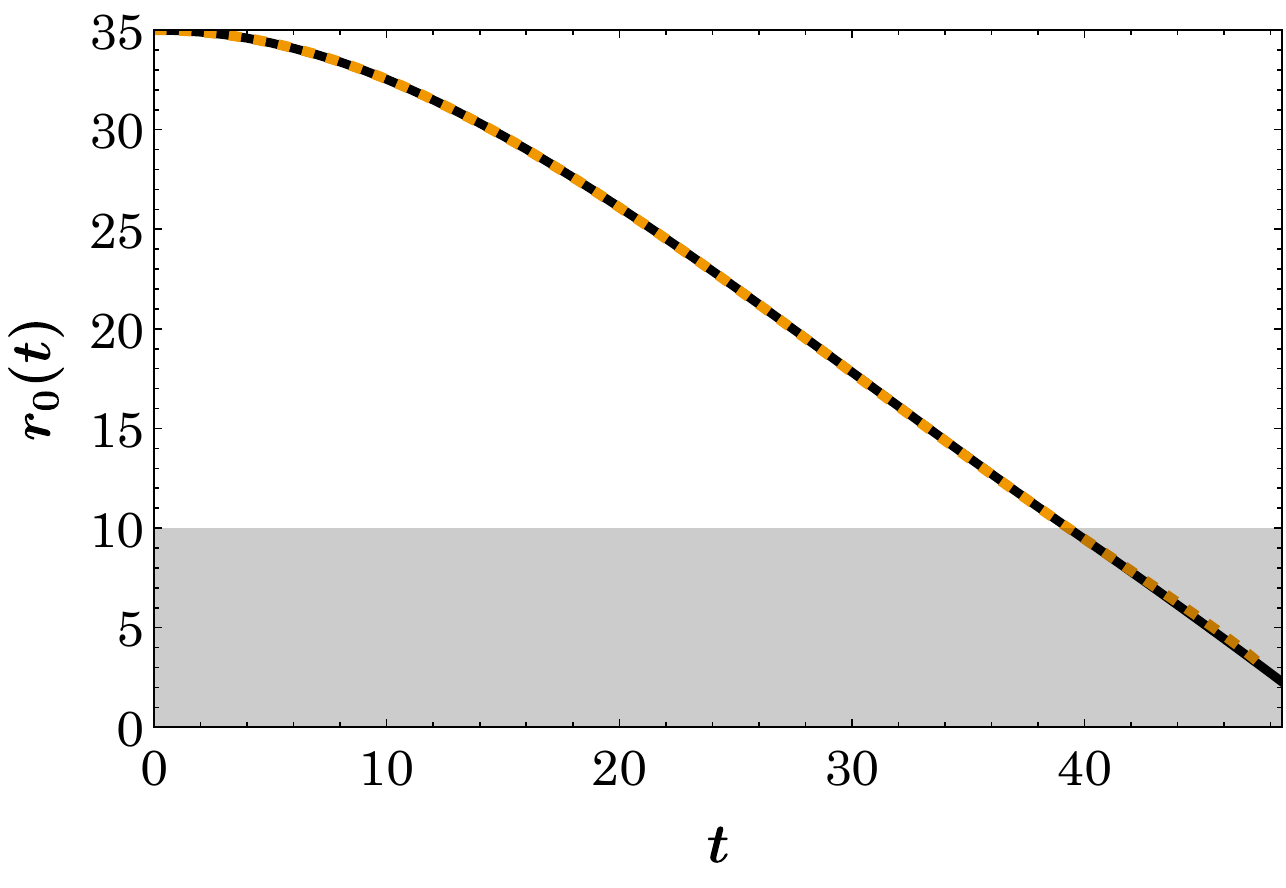}}}
    \qquad
    \subfloat{{\includegraphics[height=5cm]{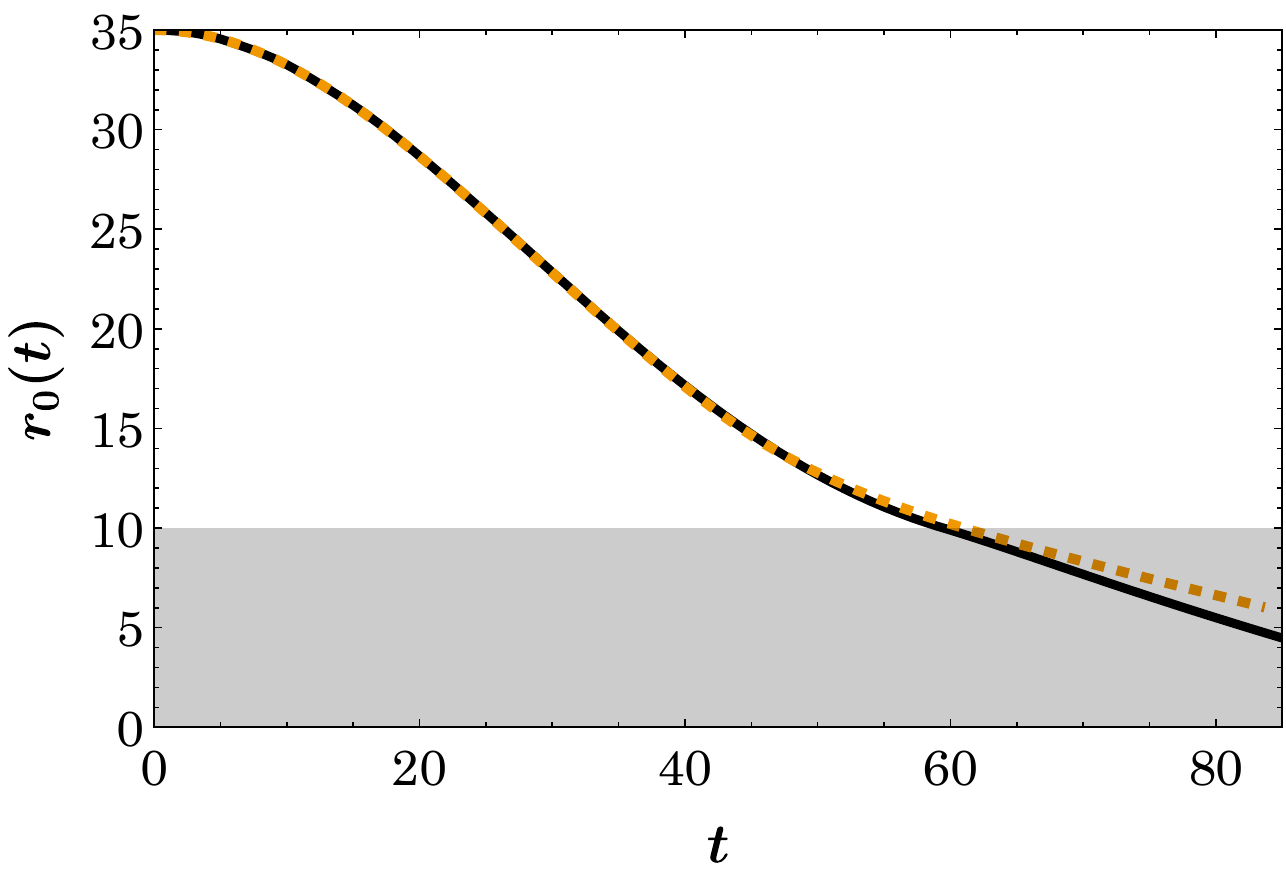}}}
    \caption{Comparison of the position of the center of the kink of the solutions of the full model (black solid line) with the approximate model \eqref{eq33a}, \eqref{eq33b} with the new ansatz \eqref{zeta_3} (yellow - dashed  line). In this case,  $R=10$ (see gray area that represent the star), $r_0 = 35$ and $\Dot{r}_0 = 0$,  $\gamma(t=0) = 1$, on the left $M=1$ and on the right $M=4$.}
    \label{fig_11}
\end{figure}

The resulting equations of motion that describe this system are in full analogy to the system of equations \eqref{eq3a} and \eqref{eq3b}
\begin{equation}
\label{eq33a}
   {\cal M} \, \Ddot{r}_0 + \Tilde{b} ~\Ddot{\gamma} +\frac{1}{2} \,  \frac{\partial {\cal M}}{\partial r_0} \, \Dot{r}_0^2  + \frac{\partial {\cal M}}{\partial \gamma} \, \Dot{\gamma} ~\Dot{r}_0  + \left( \frac{\partial \Tilde{b}}{\partial \gamma}
   - \frac{1}{2} \frac{\partial \Tilde{a}}{\partial r_0}\right) \Dot{\gamma}^2 + \frac{\partial V}{\partial r_0} = 0 ,
\end{equation}
and
\begin{equation}
\label{eq33b}
 \Tilde{a} \, \Ddot{\gamma}  +
 \Tilde{b} \, \Ddot{r}_0  +
 \frac{1}{2} \frac{\partial \Tilde{a}}{\partial \gamma} \, \Dot{\gamma}^2  + 
 \frac{\partial \Tilde{a}}{\partial r_0} \, \Dot{\gamma} \, \Dot{r}_0  + 
 \left( 
 \frac{\partial \Tilde{b}}{\partial r_0}
   - \frac{1}{2} \frac{\partial {\cal M}}{\partial \gamma} 
   \right) \Dot{r}_0^2 + 
   \frac{\partial V}{\partial \gamma} = 0 .
\end{equation}
The course of the collapse 
i.e. shrinking of the kink, up to the radius of $r_0=0.$
is shown in Figure \ref{fig_11}. This figure also includes a comparison of the trajectory obtained based on the field model (black line) with the effective model based on the equations \eqref{eq33a} and \eqref{eq33b}  (yellow dotted line). Note that the agreement of two approaches is very good both for mass $M=1$ (left panel) and for mass $M=4$ (right panel). In particular, the agreement for $M=1$ is striking. 
The radius of the star is equal to $R=10$. In the figure, the gray area represents the location of the star. In the simulations, the wall initially rested at a distance $r_0=35$ from the origin of the coordinate system. 
Summarizing, it turns out that the behavior of a system with spherical symmetry in $3+1$ dimensions is not significantly different from that of a system in $2+1$ dimensions with rotational symmetry. In particular, the energy landscapes in both cases are very similar (see Fig. \ref{fig_03} and \ref{fig_10}). In both cases, the presence of a gravitating object inside the wall is unable to stop its collapse.

\section{Conclusions and Future Challenges}
The present work focuses on two issues. The first is the potential of stopping the collapse of a domain wall surrounding a massive object,
given the existence of a repulsive force between the domain wall and normal matter \cite{Ipser1984}. 
Our simulations show that such a stabilization does not occur, and the domain wall inevitably collapses transforming into a dominant vacuum state, even though this effect is delayed by the presence of the
massive object (neutron star)  \cite{animation}. 

The second issue is the possibility of an effective description of the kink front in the presence of a non-trivial gravitational background. It is known that if the gravitational fields are weak (the masses of the sources are small) and accompanied by slow motions of the front, then the standard effective description can be obtained by the two ansatz equations \eqref{phi_k} and \eqref{zeta_1}. For spherical symmetry, the only dynamical variable of the effective model is the radial position of the kink.  This standard approach leads to relatively correct results for very small masses. The situation deteriorates significantly if we are dealing with a strong gravitational field. 
Indeed, we have illustrated how inadequate this approach is in
such a case.

To amend this discrepancy, three effects must be taken into account. The first is related to including the effect of the gravitational field on the kink at its current location.  The next two are related to dynamic and kinematic changes in the width/thickness of the kink. These effects are described by the ansatz given by the \eqref{phi_k} and \eqref{zeta_3} equations. The model created based on this ansatz contains two dynamic degrees of freedom: the position and the width of the kink.
A comparison of the trajectories obtained based on the effective model 
and the original field model shows a striking agreement between the two predictions.
This compatibility is all the more valuable for the 
non-perturbative regime of the larger masses that we deal with. 
It is also worth noting that compatibility also occurs at times when the deviation of the shape of the field configuration from that of the ansatz begins to become apparent. Obviously, the end of the  evolution cannot be described with an effective model because, due to its shrinking towards
a vanishing radius, the kink configuration transforms into a vacuum configuration which is not captured by the ansatz.

The decay process of the kink configuration was studied numerically, and it leads to the only stable configuration in this situation, which is the vacuum state. The vacuum itself in the presence of a gravitational field has a rather complex spectrum of excitations, however, all the excitations reflect its stability. Here, we have
shown how the presence of the neutron star has a fairly minimal
role (predominantly in the form of a time-shift) and how the
principally linear dynamics is sufficient to appreciate the
qualitative evolution of the relevant process.

A natural next step for further research is to apply a similar ansatz to the case of a flat (planar) kink front. Due to the lack of rotational symmetry of the initial conditions and boundary conditions in this case, this leads to the need to leverage a dynamical variable describing the position of the front dependent on the time coordinate, but also on the pseudo-Cartesian coordinate directed along the kink front, e.g., a center variable
such as $Y=Y(X,t)$, in a way similar to what was discussed
in~\cite{Kevrekidis2018}. A similar situation will occur for the dynamic variable describing the thickness of the kink front. This will allow the model to vary both the position and thickness of the kink depending on the position of the selected point along the front. This will naturally permit to describe, among other things, the difference in the strength of gravitational interactions to which different parts of the kink front may be subjected.
Developing such models both in two- and in three-dimensional
settings and exploring their potential for describing such kinks
in higher dimensions,
even more so in the context of azimuthal (or more generally 
transverse) perturbations is an interesting topic for future
studies.

In this paper, we considered the motion of a rotationally symmetric domain wall in a Schwarzschild-type geometry without singularities. Because of this assumption, we limited ourselves to the parameter $M/R < 4/9$. Fulfilling this condition makes both the metric outside the star and the metric inside the star regular. The situation changes when we enter the regime $4/9<M/R<1/2$. It turns out that in this area outside the gravitating object the singularity does not occur. The situation is different with the internal metric, where the singularity appears. To be precise, it takes the form of a circle (in the 2+1 dimensional case), or a sphere (in the 3+1) dimensional case. The radius of this singular sphere increases, with the increase of the parameter $M/R$, from zero for $M/R=4/9$ to $r=R$ for $M/R=1/2$. An important property of the mentioned singularity is that we are faced with, not only the singularity of the metric, but also the singularity of the Ricci curvature scalar. Due to the existing singularity, the interpretation of the course of simulations and their reliability is quite problematic. In any case, we found that if the initially spherical domain wall is resting and has a radius larger than the radius of the star then, upon approaching the singularity, the wall is compressed (its thickness gradually reduces to zero) and the wall itself stops at the singularity. On the other hand, if the initial radius of the domain wall is smaller than the radius of the singularity sphere then the initially resting wall collapses into the vacuum. The singularity itself in the intrinsic metric needs to be examined more closely, in particular the existence of an event horizon for it, i.e. whether it is not a naked singularity.

\section*{Appendix A}
In order to avoid singularities on the surface of $r=R$ in the derivatives of the functions $A$, $B$ and $G$, we replace the step function by hyperbolic tangents i.e.
$$\Theta(r-R) \rightarrow \frac{1}{2}\,(\tanh(a(r - R)) + 1) \,\,\,\mathrm{and}\,\,\, \Theta(R-r) \rightarrow \frac{1}{2}\,(\tanh(a(R-r)) + 1)$$
with $a=5$. The next three figures compare the expressions for the $A$, $B$ and $G$ functions written using the $\Theta(r-R), \, \Theta(R-r)$ functions (solid black line) and those written using the $\tanh(r-R), \, \tanh(R-r)$ functions (dashed red line). The gray area corresponds to a star with radius $r=R$. It can be seen that the numerical compliance is very good.
\begin{figure}[!ht]
    \centering
    \subfloat{{\includegraphics[height=4.5cm]{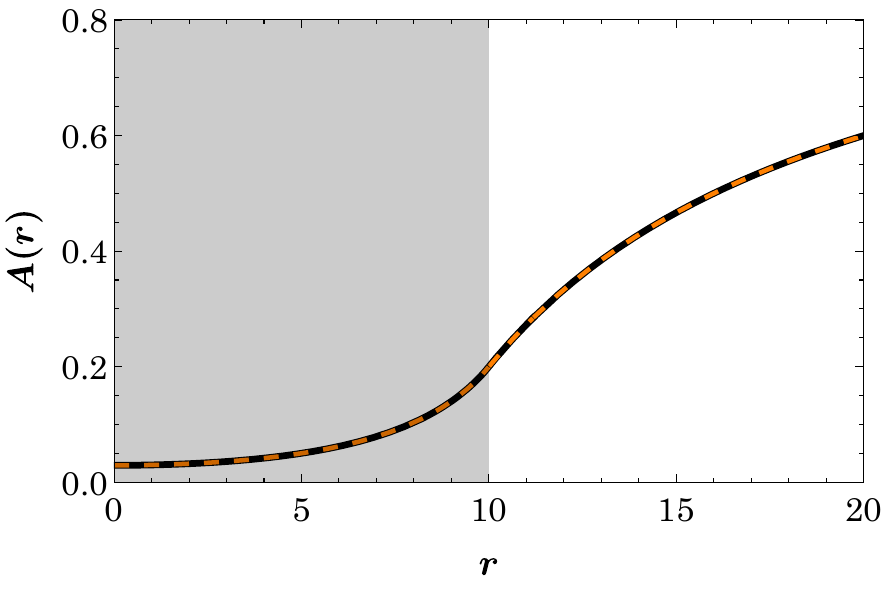}}}
    \quad
    \subfloat{{\includegraphics[height=4.5cm]{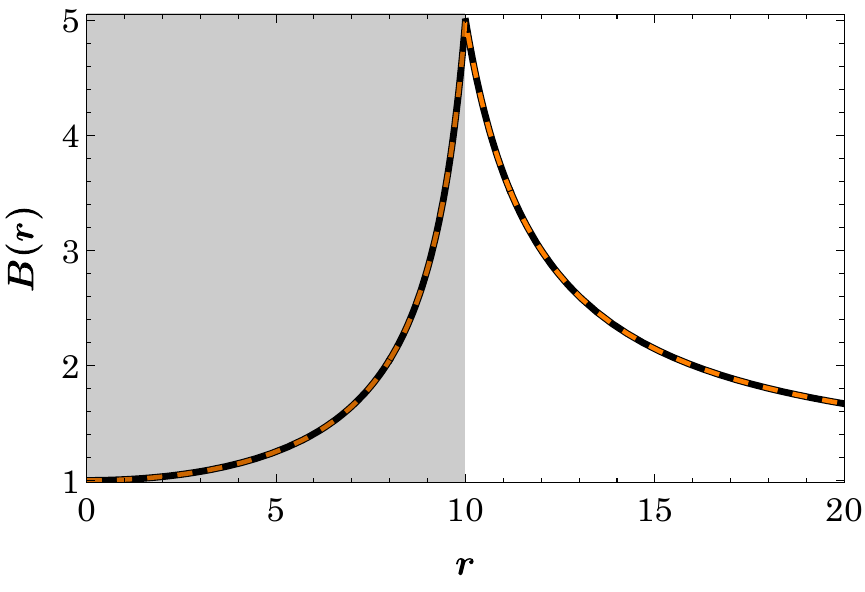}}}
    \quad
    \subfloat{{\includegraphics[height=4.5cm]{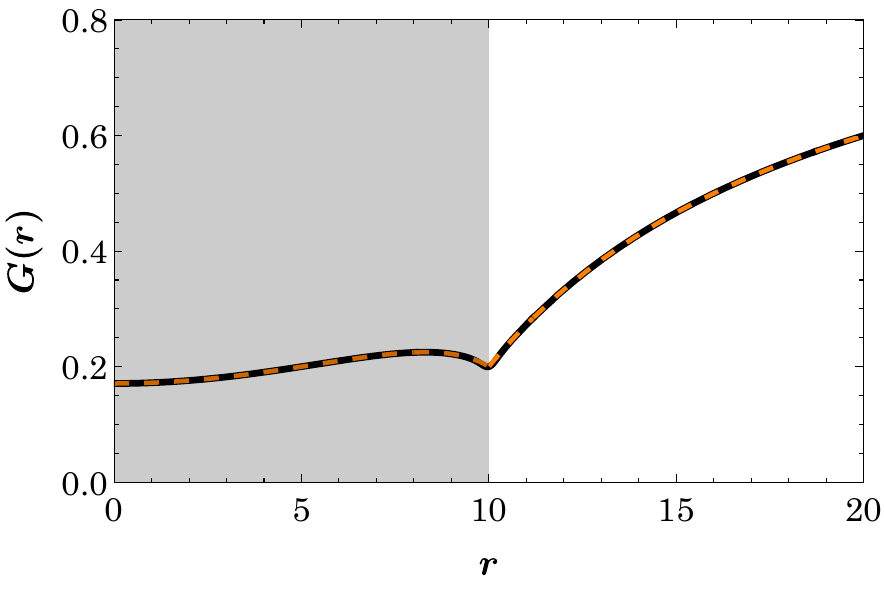}}}
    \caption{Functions $A(r)$, $B(r)$  defining the  components of the metric  tensor and the auxiliary function  $G(r)$ for $M=4$ and $R=10$. The white area is outside the star, and the gray area is inside.}
    \label{fig_12}
\end{figure}

\section*{Appendix B}
In this supplement, definitions of all coefficients present in subsection 3.2 are listed.
In the case of the non-relativistic ansatz described in subsection 3.2.1 and given by the function $ \zeta = \frac{1}{\sqrt{2}}  \,  (r - r_0)$, the coefficients are of the form 
\begin{equation}
    {\cal M}_1(r_0) \equiv  \pi  \int_0^{\infty} r \sqrt{\frac{B(r)}{A(r)}} \sech^4 \zeta \,
    \, d r ,
\end{equation}
\begin{equation}
    V_1(r_0) \equiv \frac{1}{2} \, \pi  \int_0^{\infty} r \sqrt{\frac{B(r)}{A(r)}} \sech^4 \zeta 
     \left( A(r) +  \frac{A(r)}{B(r)}  \right)
    \, d r .
\end{equation}
The second ansatz explicitly takes into account the interaction of the gravitational field with the kink front. The ansatz is described in subsection 3.2.2 and is given by function $ \zeta = \frac{1}{\sqrt{2}} \, \sqrt{B(r_0)} \,  (r - r_0)$.  The coefficients in this case are as follows 
\begin{equation}
    {\cal M}_2(r_0) \equiv 2 \pi  \int_0^{\infty} r \sqrt{\frac{B(r)}{A(r)}} \sech^4 \zeta 
    \left( \frac{1}{2 B(r_0)} \frac{\partial B(r_0)}{\partial r_0} \, \zeta - \sqrt{\frac{B(r_0)}{2}} \right)^2
    \, d r ,
\end{equation}
\begin{equation}
    V_2(r_0) \equiv \frac{1}{2} \, \pi  \int_0^{\infty} r \sqrt{\frac{B(r)}{A(r)}} \sech^4 \zeta 
    \left( A(r) + A(r) \frac{B(r_0)}{B(r)}  \right)
    \, d r .
\end{equation}
The third ansatz explicitly takes into account not only the interaction of the gravitational field with the kink front but also the effect of the field on the thickness of the kink. The ansatz is described in subsection 3.2.3 and is given by the function $ \zeta = \frac{1}{\sqrt{2}} \, \sqrt{B(r_0)} \, \gamma (r - r_0)$.  In this case, we have two dynamical variables $r_0=r_0(t)$ and $\gamma=\gamma(t)$. The coefficients in this case take the form
\begin{equation}
    {\cal M}_3(r_0,\gamma) \equiv 2 \pi  \int_0^{\infty} r \sqrt{\frac{B(r)}{A(r)}} \sech^4 \zeta 
    \left( \frac{1}{2 B(r_0)} \frac{\partial B}{\partial r_0} \, \zeta - \sqrt{\frac{B(r_0)}{2}} \,\, \gamma \right)^2
    \, d r ,
\end{equation}
\begin{equation}
    V_3(r_0,\gamma) \equiv \frac{1}{2} \, \pi  \int_0^{\infty} r \sqrt{\frac{B(r)}{A(r)}} \sech^4 \zeta 
    \left( A(r) + A(r) \frac{B(r_0)}{B(r)} \, \gamma^2 \right)
    \, d r ,
\end{equation}
\begin{equation}
    a(r_0,\gamma) \equiv 2 \pi  \int_0^{\infty} r \sqrt{\frac{B(r)}{A(r)}} \sech^4 \zeta \,\,
   \frac{\zeta^2}{\gamma^2} 
    \, d r ,
\end{equation}
\begin{equation}
    b(r_0,\gamma) \equiv 2 \pi  \int_0^{\infty} r \sqrt{\frac{B(r)}{A(r)}} \sech^4 \zeta 
    \left( \frac{1}{2 B(r_0)} \frac{\partial B(r_0)}{\partial r_0} \, \zeta - \sqrt{\frac{B(r_0)}{2}} \,\, \gamma \right) \, \frac{\zeta}{\gamma}
    \, d r .
\end{equation}

\section*{Appendix C}
The model describing the domain wall in $3+1$ dimensions is fully analogous to the last model describing a similar system in $2+1$ dimensions. As before, it describes the dynamics of the system using two degrees of freedom $r_0=r_0(t)$ and $\gamma=\gamma(t)$. The coefficients appearing in subsection 4.2 and describing the dynamics of the spherical domain wall are defined by the following formulas
\begin{equation}
    {\cal M}(r_0,\gamma) \equiv 4 \pi  \int_0^{\infty} r^2 \sqrt{\frac{B(r)}{A(r)}} \sech^4 \zeta 
    \left( \frac{1}{2 B(r_0)} \frac{\partial B}{\partial r_0} \, \zeta - \sqrt{\frac{B(r_0)}{2}} \,\, \gamma \right)^2
    \, d r ,
\end{equation}
\begin{equation}
    V(r_0,\gamma) \equiv  \pi  \int_0^{\infty} r^2 \sqrt{\frac{B(r)}{A(r)}} \sech^4 \zeta 
    \left( A(r) + A(r) \frac{B(r_0)}{B(r)} \, \gamma^2 \right)
    \, d r ,
\end{equation}
\begin{equation}
    \Tilde{a}(r_0,\gamma) \equiv 4 \pi  \int_0^{\infty} r^2 \sqrt{\frac{B(r)}{A(r)}} \sech^4 \zeta \,\,
   \frac{\zeta^2}{\gamma^2} 
    \, d r ,
\end{equation}
\begin{equation}
    \Tilde{b}(r_0,\gamma) \equiv 4 \pi  \int_0^{\infty} r^2 \sqrt{\frac{B(r)}{A(r)}} \sech^4 \zeta 
    \left( \frac{1}{2 B(r_0)} \frac{\partial B(r_0)}{\partial r_0} \, \zeta - \sqrt{\frac{B(r_0)}{2}} \,\, \gamma \right) \, \frac{\zeta}{\gamma}
    \, d r .
\end{equation}

\section*{Acknowledgments} 
This material is based upon work supported by the U.S. National Science Foundation under the awards PHY-2110030 and DMS-2204702 (PGK). The work at AGH University was supported by the National Science Centre, Poland, Grant OPUS: 2021/41/B/ST3/03454, and the “Excellence Initiative-Research University” program for AGH University of Krakow (JG).

\printbibliography

\end{document}